\relax
\documentclass[letterpaper]{article} 
\usepackage{aaai22}  
\usepackage{times}  
\usepackage{helvet}  
\usepackage{courier}  
\usepackage[hyphens]{url}  
\usepackage{graphicx} 
\usepackage{natbib}  
\usepackage{caption} 
\DeclareCaptionStyle{ruled}{labelfont=normalfont,labelsep=colon,strut=off} 
\usepackage{amsmath}
\usepackage{booktabs}
\usepackage{multirow}
\usepackage{amsfonts,amssymb}
\usepackage{soul}
\usepackage[
 font=normalsize 
 ]{subfig}

\usepackage[dvipsnames]{xcolor}
\definecolor{zlg}{rgb}{0.858, 0.188, 0.478}
\title{View Blind-spot as Inpainting: Self-Supervised Denoising with Mask Guided Residual Convolution}
\author {
    Yuhongze Zhou,\textsuperscript{\rm 1}
    Liguang Zhou, \textsuperscript{\rm 2,3}
    Tin Lun Lam, \textsuperscript{\rm 2,3} \thanks{Corresponding Author}
    Yangsheng Xu \textsuperscript{\rm 2,3}
}
\affiliations {
    \textsuperscript{\rm 1} McGill University\\
    \textsuperscript{\rm 2} The Chinese University of Hong Kong, Shenzhen\\
    \textsuperscript{\rm 3} Shenzhen Institute of Artificial Intelligence and Robotics for Society\\
    yuhongze.zhou@mail.mcgill.ca, \{liguangzhou@link., tllam@, ysxu@\}cuhk.edu.cn
}

\usepackage{bibentry}

\begin{document}

\maketitle

\begin{abstract}
In recent years, self-supervised denoising methods have shown impressive performance, which circumvent painstaking collection procedure of noisy-clean image pairs in supervised denoising methods and boost denoising applicability in real world. One of well-known self-supervised denoising strategies is the blind-spot training scheme. However, a few works attempt to improve blind-spot based self-denoiser in the aspect of network architecture. In this paper, we take an intuitive view of blind-spot strategy and consider its process of using neighbor pixels to predict manipulated pixels as an inpainting process. Therefore, we propose a novel Mask Guided Residual Convolution (MGRConv) into common convolutional neural networks, e.g. U-Net, to promote blind-spot based denoising. Our MGRConv can be regarded as soft partial convolution and find a trade-off among partial convolution, learnable attention maps, and gated convolution. It enables dynamic mask learning with appropriate mask constrain. Different from partial convolution and gated convolution, it provides moderate freedom for network learning. It also avoids leveraging external learnable parameters for mask activation, unlike learnable attention maps. The experiments show that our proposed plug-and-play MGRConv can assist blind-spot based denoising network to reach promising results on both existing single-image based and dataset-based methods. 
\end{abstract}

\section{Introduction}
Image denoising is one of the most fundamental tasks in image restoration tasks. A noisy image $y$ can be modeled as
\begin{equation}
y=x+n, 
\end{equation}
where $x$ is a clean image, and $n$ is random noise. In recent years, with deep learning flourishing in computer vision area, the performance of supervised denoising methods, e.g. U-Net~\cite{ronneberger2015u}, RED-Net~\cite{mao2016image}, DnCNN~\cite{zhang2017beyond}, MemNet~\cite{tai2017memnet}, SGN~\cite{gu2019self}, MIRNet~\cite{zamir2020learning}, MPRNet~\cite{zamir2021multi} and IPT~\cite{chen2021pre} have greatly surpassed traditional approaches. However, model trained by synthetic noisy images is hard to generalize to realistic noisy images, and requires collecting sufficient real-world noisy-clean image pairs, which is challenging and involves heavy labor resources. To alleviate the aforementioned problems, unsupervised and self-supervised methods, using only noisy images, have sprung up.

On one hand, a brief recap for these methods can be: 1) use multiple noisy images for training: it can be two different noisy observations of the same scene~\cite{lehtinen2018noise2noise}, or noisier-noisy image pairs, the noisier one of which originates from noisy image by adding synthesized noise~\cite{moran2020noisier2noise,xu2020noisy}, or noisy image pairs generated by random neighbor sub-sampler~\cite{huang2021neighbor2neighbor}. 2) introduce blind-spot training scheme: it can be to manipulate noisy images via randomly masking out/replacing pixels and calculate loss function on manipulated region~\cite{krull2019noise2void,batson2019noise2self,quan2020self2self}; it can be a novel network architecture incorporating with noise modeling to further boost performance~\cite{laine2019high,wu2020unpaired}. It is noted that, with the passage of research progress, denoisers using multiple noisy images has been upgraded from requirement of multiple noisy observations of the same scene~\cite{lehtinen2018noise2noise} to noisy image generation by smart random sub-sampling~\cite{huang2021neighbor2neighbor}; blind-spot training scheme from $l_2$ (MSE) masking loss~\cite{krull2019noise2void,batson2019noise2self,quan2020self2self} to novel network with noise modeling~\cite{laine2019high,wu2020unpaired}. \textbf{Nevertheless, relatively few works attempt to propose more efficient network module for MSE masking loss training, although Self2Self~\cite{quan2020self2self} has introduced partial convolution~\cite{liu2018image} into denoising network.}

On the other hand, apart from the classification of previous quick review, denoising without clean images can also be roughly categorized into two domains by the magnitude of training data, i.e. dataset-based and single-image based training. Here we focus on blind-spot based denoiser. The dataset-based denoising approaches boosted by novel network and detailed noise modeling~\cite{laine2019high,wu2020unpaired} are time-efficient in inference and have shown impressive performance. Based on blind-spot training scheme under the assumption of zero-mean and i.i.d. noise, single-image based denoisers~\cite{ulyanov2018deep,krull2019noise2void,batson2019noise2self,quan2020self2self} require longer denoising time and heavier computational resources and cannot handle Poisson noise well. It is because that the noise modeling way is too simplistic to tackle various noise distribution, or/and one single image contains deficient information, or/and randomly manipulating noisy image and calculating loss only on manipulated/mask-out pixels blind the network to see the whole image in each back-propagation. Usually, following blind-spot strategy, single-image based denoiser is a special case of dataset-based denoiser, which means that theoretically these two can be transferred between each other. Unfortunately, the dropout-based sampling generation of Self2Self~\cite{quan2020self2self} blocks this interconvertible path. Methods proposed by Laine et al.~\cite{laine2019high} and Wu et al.~\cite{wu2020unpaired} are hardly extended to single-image based. The single image version of Noise2Void~\cite{krull2019noise2void} and Noise2Self\cite{batson2019noise2self} has performance concession compared to dataset-based, while Noise2Self focuses on monochromatic images and cannot handle color images well. \textbf{Therefore, most denoising methods have their own drawbacks due to their noise modeling and training strategy (not network structure), which are unlikely to be circumvented.}

\textbf{Thus, how can we propose a more effective network structure to promote both single-image based and dataset-based denoisers under blind-spot scheme and remedy performance concession caused by noise modeling and training strategy?}

Intending to introduce a more robust network for self-denoising and taking the one less traveled by, we provide a different and intuitive perspective on the principle behind masking-based blind-spot training scheme~\cite{krull2019noise2void,batson2019noise2self,quan2020self2self}. We consider the procedure of leveraging neighbor pixels to predict random mask-out uncertain regions as an inpainting process. To the best of our knowledge, we are the first to consider self-supervised denoising in this kind of perspective. We propose Mask Guided Residual Convolution (MGRConv) to boost self-denoising in an inpainting manner, which fits the blind-spot masking strategy well. With more stable and light-weight training compared to learnable attention maps~\cite{xie2019image}, the MGRConv is equipped with a kindly-confined mask learning strategy for dynamic information gating, different from hard clipping of partial convolution~\cite{liu2018image} and freeform training of gated convolution. For training, the random masking strategy of noisy image simulates variations of input noisy image and enables network to estimate mean image from those noisy images as clean image.

In this paper, we show that, with the assistance of a carefully-designed plug-and-play MGRConv, blind-spot based denoiser in both single-image and dataset-based training aspects can have more potential in both denoising performance and research prospects. We validate our MGRConv by a series of experiments on both synthetic and real-world noisy images. Extensive experiments show that our proposed network module can facilitate self-denoising performance convincingly. 





\section{Related Work}
In the last few years, supervised image denoising~\cite{ronneberger2015u,mao2016image,zhang2017beyond,tai2017memnet,zhang2018ffdnet,lefkimmiatis2018universal,plotz2018neural,guo2019toward,gu2019self,zamir2020learning,zamir2021multi,chen2021pre} has achieved startling performance. However, there still exists a gap between synthesized noisy-clean image pairs and realistic noisy images. To bridge this gap, tremendous real-captured aligned noisy-clean image pairs are required, whose collection is challenging and painstaking.

To circumvent limitations of supervised methods, unsupervised/self-supervised denoising using only noisy images has been well investigated, which can be categorized into two groups, i.e. non-learning and learning methods. Non-learning methods include BM3D~\cite{dabov2007image}, NLM~\cite{buades2005non}, and WNNM~\cite{gu2014weighted}. Learning-based approaches can be separated into two branches according to the magnitude of training data, i.e. dataset-based and single-image based.

For dataset-based denoising, current approaches can be roughly divided into the following parts by methodology. \textbf{1) more than one noisy images:} Noise2Noise~\cite{lehtinen2018noise2noise} trains a denoiser with pairs of two different noisy observations of the same clean image, whose performance is close to supervised denoising. Furthermore, Noisier2Noise~\cite{moran2020noisier2noise} uses noisier-noisy image pairs to handle white noise and spatially correlated noise. Xu et al.~\cite{xu2020noisy} propose Noisy-as-Clean (NAC) strategy that utilizes corrupted images and synthetic images containing original corruption and another similar corruption, to train self-supervised denoising networks. Recently, Neighbor2Neighbor~\cite{huang2021neighbor2neighbor} proposes a random neighbor sub-sampler for noisy image pair generation and a regularizer as additional loss for better performance. \textbf{2) blind-spot:} Noise2Void (N2V)~\cite{krull2019noise2void} introduces blind-spot mechanism to avoid learning identity mapping by excluding the pixel itself from the receptive field of each pixel. Noise2Self (N2S)~\cite{batson2019noise2self} and Probabilistic Noise2Void (PN2V)~\cite{krull2020probabilistic} also follow similar training scheme. Laine et al.~\cite{laine2019high} introduce a novel neural network structure to build blind-spot inside CNN that combines multiple branches which have their half-plane receptive field but exclude the center pixel. Wu et al.~\cite{wu2020unpaired} introduce Dilated Blind-Spot Network (DBSN) to incorporate self-supervised learning and knowledge distillation. One limitation of blind-spot training scheme is whole image information loss in each back-propagation, which somehow slowdowns the training procedure.

\begin{figure*}[thpb]
\centering
\includegraphics[scale=0.5]{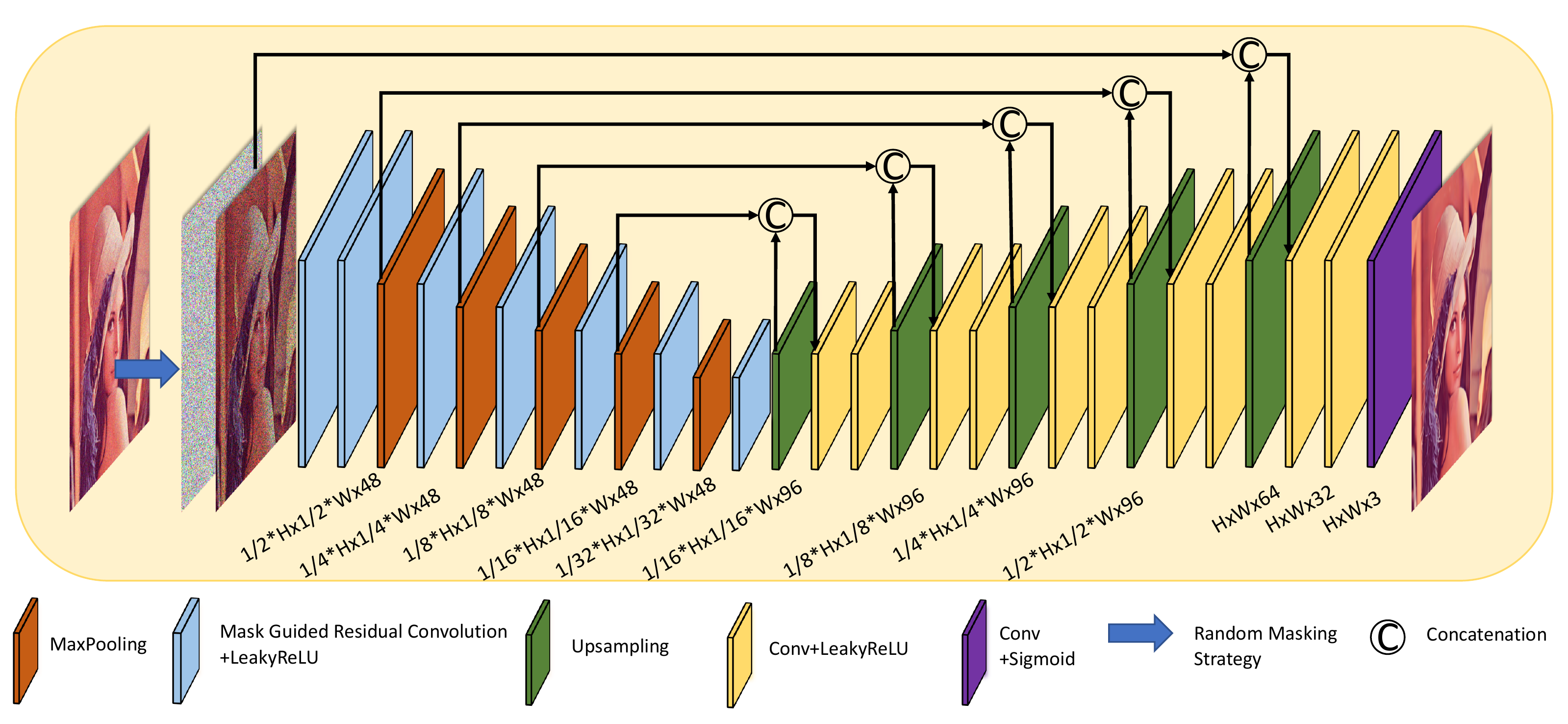}
\caption{The overview of our denoising network}
\label{fig:denoising_net}
\end{figure*}
For single-image based denoiser whose training set contains only one noisy image, it can be considered as a special case of dataset-based one. Deep Image Prior (DIP)~\cite{ulyanov2018deep} employs a generative network to capture image statistics prior and map a random noise to a denoised image by early stopping, but its result is severely affected by training iterations. Self2Self~\cite{quan2020self2self} generates Bernoulli-sampled instances to cater for blind-spot scheme and reduces the variance of MSE loss by dropout, which promotes denoising network performance significantly. The above-mentioned N2V and N2S can also be extended to single-image based denoiser.

Here comes the summary. Dataset-based denoisers, especially explicit noise distribution modeling, have reached impressive performance but will degrade greatly in real applications if noise distribution is unknown. For single-image based denoiser, it is usually based on zero-mean and i.i.d. noise assumption, which can be more flexible in practice but more time-consuming and sometimes a bit inferior than dataset-based methods.

Different from past researches, in this paper, we commit to introducing a novel network module, MGRConv, to boost, unite, and bridge single-image based and dataset-based denoisers under blind-spot scheme in an inpainting manner and compensate performance degradation caused by their own noise modeling and training strategy.


\section{Approach}
In this section, we present image formulation, demonstrate our motivation, introduce our MGRConv by revisiting previous inpainting convolutions, and then introduce the overview of our denoising network in Fig.~\ref{fig:denoising_net}.
\subsection{Image Formulation}
Consider that denoising is to estimate the clean image $x$ from noisy image $y$, $y=s+n$, with the unobserved clean image as $s$ and noise denoted as $n$. Assume pixels in $s$ are not independent; on the contrary, pixel $s_i$ depends on the context of its neighboring pixels $\Omega_{y_i}$, which corresponds to the receptive field sans the pixel $y_i$ itself in convolutional neural networks. Further, noise is assumed to be zero-mean, i.e. $\mathbb{E}(n)=0$, independent between each other, and independent of clean image context. Then, we can obtain $\mathbb{E}(y)=s$. Therefore, we can know that if training neural network with various images $y$ with the same signal $s$ but different noise realizations $n$, the network output $\mathbb{E}(x|\Omega_y)$, the mean over all possible clean pixels given the neighboring context, will be near to the result of a supervised denoising regression model with $l_2$ loss that estimates $\mathbb{E}(x|y,\Omega_y)$, the mean over all possible clean pixels given the noisy pixel and its neighboring context.

Based on the above theory, blind-spot based denoising schemes, e.g. Noise2Void~\cite{krull2019noise2void}, Noise2Self~\cite{batson2019noise2self}, and Self2Self~\cite{quan2020self2self}, have been proposed to reach self-denoising without additional post-processing, e.g. Laine et al~\cite{laine2019high} that considers the whole corrupted image $y$ during test time.

\subsection{Motivation}
From a theoretic perspective, assuming that noise is zero-mean and independent among pixels, blind-spot based denoising schemes~\cite{quan2020self2self,laine2019high} not only use the surrounding context to predict masked pixels to avoid identity mapping but also provide different variants of single noisy image for better estimation of the expectation of MSE between clean and noisy image. From a more straightforward and intuitive view, the process of generating training data by blind-spot masking strategy for network to predict clean data in certain area with noisy neighborhood provided is similar to an inpainting task that clean pixels are inferred from surrounding valid pixels. For example, Self2Self masking strategy~\cite{quan2020self2self} can be seen as a more freeform way of blind-spot network compared to replacing randomly picked pixels with random neighborhood pixels~\cite{krull2019noise2void}. Each Bernoulli sampled noisy image generated by random dropout feeds into a network whose loss is calculated only on the area that is masked out by dropout. Therefore, we argue that a more task-adaptive neural network should definitely promote denoising performance.

With motivation stated above, we model blind-spot based denoising procedure as an inpainting problem and introduce a novel mask convolution to further improve performance. Let us revisit inpainting convolutions in literature first.
\subsection{Revisiting Inpainting Convolutions}
\subsubsection{Partial Convolution (PConv)} Partial convolution~\cite{liu2018image} consists of three steps, i.e. mask convolution, feature re-normalization, and mask updating, and can be formulated as follows:
\begin{equation}
I^{\prime}=
\begin{cases}
\sum{\sum{{W} \cdot ({X}\odot {M}) \frac{sum(\mathbf{1})}{sum(M)}}}, \text{if sum(M)}>0,\\
0, \qquad \qquad \qquad \qquad \qquad \quad\text{otherwise},
\end{cases},
\end{equation}
where $I^{\prime}$ is the updated image feature, $W$ are convolutional filters, $X$ is image feature for the current convolution, $M$ is the corresponding mask, $\odot$ denotes element-wise multiplication, $\mathbf{1}$ has the same shape as $M$ but with all elements being 1. After partial convolution operation, the convolved mask is set to 1 if sum(M)$>$0, otherwise 0. Partial convolution has shown remarkable performance on inpainting task and Self2Self denoising network. However, 1) Partial convolution sets updated mask pixels as one at maximum that treats different neighbor pixels indiscriminately; 2) Feature hard-gating makes network learn by following handcrafted rule, which might circumvent performance.
\subsubsection{Learnable Attention Maps (LBAM)} Learnable attention maps~\cite{xie2019image} introduce an asymmetric Gaussian-shaped activation function for mask activation instead of hard gating,
\begin{equation}
\label{e1}
{M^c}=\sum{\sum{{W_M}\cdot M}},
\end{equation}
\begin{equation}
\label{e2}
{I^c}=\sum{\sum{{W_I}\cdot I}},
\end{equation}
\begin{equation}
g_A(M^c)=
\begin{cases}
a \text{exp}{(-{\gamma_l}(M^c - \mu)^2)}, \qquad \qquad M^c < \mu,\\
1+(a-1)\text{exp}{(-{\gamma_r}(M^c-\mu)^2)}, \text{otherwise},
\end{cases},
\end{equation}
\begin{equation}
I^{\prime}=\sum{\sum{{W} \cdot ({I^c}\odot {g_A(M^c)})}},
\end{equation}
and learnable mask-updating function, where $I$ is image feature, $W_M$ and $W_I$ are two different convolutional filters, $M^c$ and $I^c$ are mask and image features after convolution respectively, $a$, $\mu$, $\gamma_r$ and $\gamma_r$ are learnable parameters. The network has to learn specific parameters to model the importance distribution of pixels in , which makes sense in supervised learning but might be unstable without ground truth and also increase the training burden.
\subsubsection{Gated Convolution (GatedConv)} Gated convolution~\cite{yu2019free} fuses image and mask feature together as an aggregated feature and allows network to reweight aggregated feature by learnable gating generated by itself automatically, which produces higher-quality traditional and user-guided inpainting results. However, it may make learning lose control in a situation without ground truth.
\begin{figure}[thpb]
\centering
\includegraphics[scale=0.3]{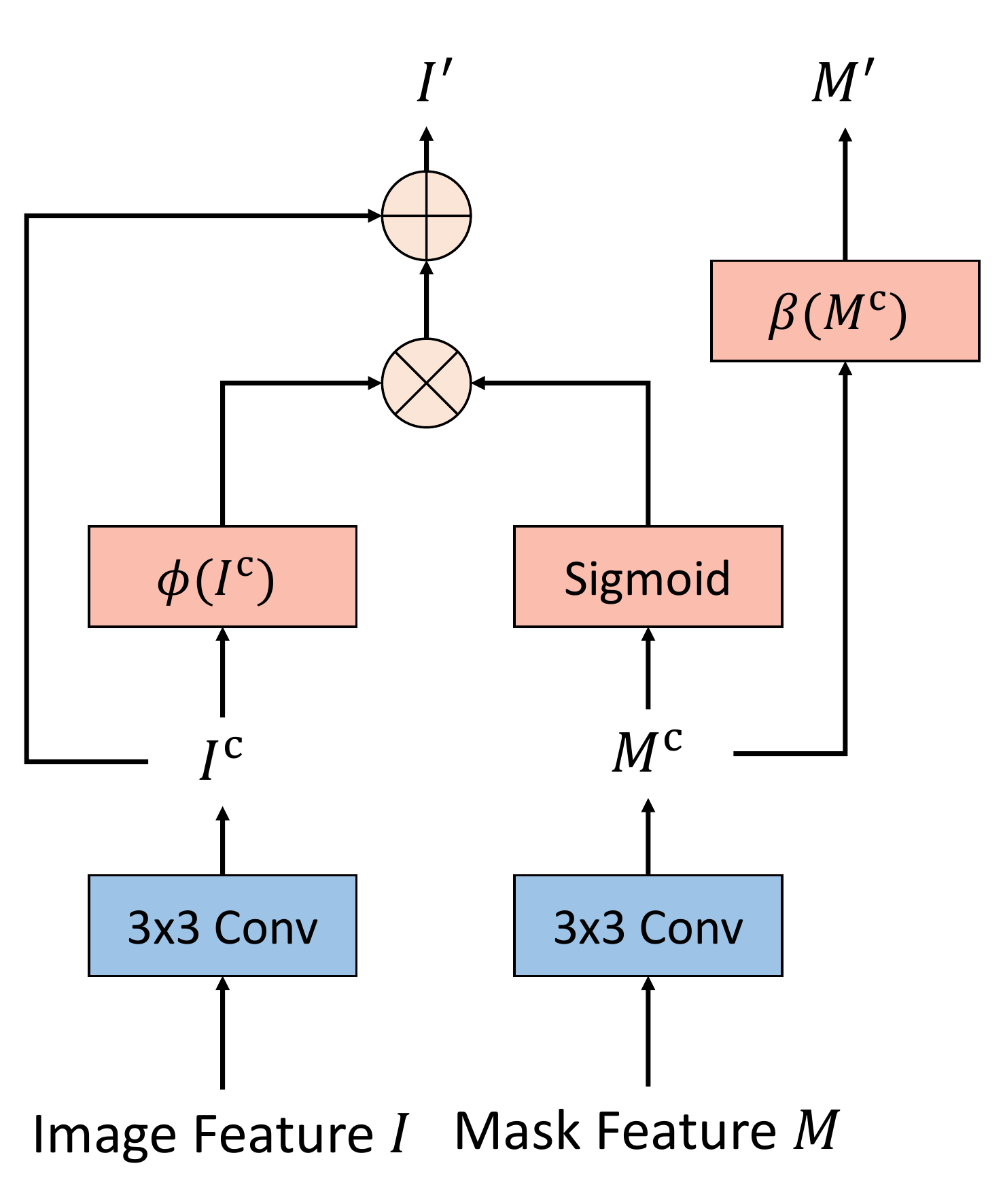}
\caption{Mask Guided Residual Convolution. ``$\bigoplus$'' means element-wise sum and ``$\bigotimes$'' denotes element-wise multiplication.}
\label{fig:mgr_conv}
\end{figure}
\subsection{Mask Guided Residual Convolution (MGRConv)}
To overcome the above limitations, we introduce our MGRConv into self-supervised denoising network. The Fig.~\ref{fig:mgr_conv} presents the procedure of MGRConv. The MGRConv first adopts equation (\ref{e1}) and (\ref{e2}). Then, the activated mask feature serves as an attention map for dynamic gating of image feature. The residual summation between gated and input image feature avoids information loss and stabilizes training. The learnable mask-updating function encourages network to progressively fill up holes of mask. The process can be represented as below:
\begin{equation}
I^{\prime}={I^c}+\phi({I^c})\odot \sigma({M^c}),
\end{equation}
\begin{equation}
{M^{\prime}}=\beta(M^c),
\end{equation}
\begin{equation}
{\beta}=(ReLU(\cdot))^{\alpha},
\end{equation}
where $M^{\prime}$ is the updated mask feature, $\phi$ can be any activation function, $\sigma$ is sigmoid function, $\beta$ is mask updating function, and $\alpha=0.8$.

The proposed MGRConv not only learns the significance weight of each pixel in each channel and fills up irregular dropout regions automatically, but also explicitly leverages mask as guidance to avoid training collapsing. It circumvents rule-based hard gating of partial convolution, makes mask updating dynamic and more flexible for learning, and can be considered as soft partial convolution. Besides, instead of explicitly modeling the importance distribution of mask as an asymmetric Gaussian-shaped function like learnable attention maps, we simplify mask activation procedure and obtains better performance without external trainable variables. To sum up, we not only find a trade-off between partial convolution~\cite{liu2018image} and learnable attention maps~\cite{xie2019image}, but also prevent playing licentiously like freeform gated convolution.

\begin{table}[thpb]
\huge
  \begin{center}
    \resizebox{\columnwidth}{!}{%
    \begin{tabular}{lccccccc}
      \toprule 
      \multirow{2}{*}{Noise Type} & \multirow{2}{*}{Methods} & \multicolumn{2}{c}{Set14} & \multicolumn{2}{c}{Kodak} & \multicolumn{2}{c}{McMaster}\\
      \cmidrule{3-8}
      &&PSNR$\uparrow$&SSIM$\uparrow$&PSNR$\uparrow$&SSIM$\uparrow$&PSNR$\uparrow$&SSIM$\uparrow$\\
      \midrule 
      &Baseline, N2N&30.51&0.913&32.39&0.945&19.60&0.575\\
      \cmidrule{2-8}
      \multirow{4}{*}{Gaussian}&DIP&29.11&0.894&29.11&0.891&30.76&0.936\\
      \multirow{4}{*}{$\sigma = 25$}&Self2Self&\underline{30.57}&\underline{0.922}&\underline{31.84}&\underline{0.940}&\underline{32.18}&\underline{\textbf{0.955}}\\
      &Noise2Void(1)&27.00&0.849&29.84&0.908&28.45&0.902\\
      \cmidrule{2-8}
      &Unet+S2S(1)&30.54&0.921&31.8&0.938&32.10&0.953\\
      &Ours+S2S(1)&\underline{\textbf{30.67}}&\underline{\textbf{0.923}}&\underline{\textbf{31.96}}&\underline{\textbf{0.941}}&\underline{\textbf{32.24}}&\underline{0.954}\\
      \bottomrule
      &Baseline, N2N&31.12&0.913&33.12&0.943&19.47&0.575\\
      \cmidrule{2-8}
      \multirow{4}{*}{Gaussian}&DIP&28.89&0.888&29.02&0.887&30.50&0.930\\
      \multirow{4}{*}{$\sigma \in [5,50]$}&Self2Self&\underline{30.98}&\underline{0.926}&\underline{32.23}&\underline{\textbf{0.938}}&\underline{32.32}&\underline{\textbf{0.956}}\\
      &Noise2Void(1)&27.23&0.865&30.06&0.905&28.21&0.897\\
      \cmidrule{2-8}
      
      &U-Net+S2S(1)&30.92&0.925&32.17&0.936&32.23&0.953\\
      &Ours+S2S(1)&\underline{\textbf{31.04}}&\underline{\textbf{0.927}}&\underline{\textbf{32.31}}&\underline{\textbf{0.938}}&\underline{\textbf{32.38}}&\underline{0.955}\\
      \bottomrule 
\end{tabular}}
\caption{Quantitative comparison (PSNR/SSIM) results of single-image based learning methods for Gaussian noise. For each noisy type, the best results in all approaches are marked in bold, and the best ones in compared approaches or our own ablation study are underlined. This highlight strategy is applied to this whole paper unless noted.}
\label{tab:denoising_single}
  \end{center}
\end{table}
\subsection{Denoising Network Structure}
The network structure is represented in Fig.~\ref{fig:denoising_net}. We adopt U-Net architecture~\cite{ronneberger2015u,quan2020self2self} and replace vanilla convolution in the encoder with our MGRConv. With a existing random masking strategy applied for training, the denoising network structure should be adapted to corresponding masking strategy. For example, in Self2Self setting, there are dropout layers inserted in each convolution layer of the decoder. In Noise2Void setting, the architecture presented in Fig.~\ref{fig:denoising_net} can be directly utilized. 


\subsection{Self-supervised Training}
In training, we adopt existing blind-spot strategies~\cite{krull2019noise2void,krull2020probabilistic}. Comprehensively speaking, the input noisy image $y$ is manipulated to generate variants of noisy instances $\hat{y}$, and the corresponding guided mask $M_{\hat{y}}$ represents positions of untouched pixels, where $M_{\hat{y}_{x,y}}=0$ when $y_{x,y}$ is manipulated, otherwise 1. The loss is measured on the manipulated area, i.e. ${(1-M_{\hat{y}})\odot(\mathcal{F_\theta}(\hat{y})-y)^2}$, where $\mathcal{F_\theta}$ represents denoising network.


\section{Experiments}
We evaluate our denoising network architecture in both single-image based and dataset-based denoising training settings by adopting existing blind-spot strategies. For single-image based setting, we adopt training schemes of both Self2Self~\cite{quan2020self2self} and Noise2Void~\cite{krull2019noise2void,krull2020probabilistic}, while we train our network with Noise2Void masking strategy without 64$\times$64 patch extraction~\cite{krull2020probabilistic} for dataset-based denoising. 
\begin{table}[thpb]
\huge
\begin{center}
\resizebox{\columnwidth}{!}{%
\begin{tabular}{lccccccc}
    \toprule
      \multirow{2}{*}{Noise Type} & \multirow{2}{*}{Methods} & \multicolumn{2}{c}{Set14} & \multicolumn{2}{c}{Kodak} & \multicolumn{2}{c}{BSD300}\\
      \cmidrule{3-8}
      &&PSNR$\uparrow$&SSIM$\uparrow$&PSNR$\uparrow$&SSIM$\uparrow$&PSNR$\uparrow$&SSIM$\uparrow$\\
      \midrule 
      &Baseline, N2N&30.51&0.913&32.39&0.945&31.15&0.938\\
      \cmidrule{2-8}
      \multirow{4}{*}{Gaussian}&Laine19-mu&29.46&0.902&30.66&0.924&28.75&0.897\\
      \multirow{4}{*}{$\sigma = 25$}&Laine19-pme&\underline{\textbf{30.73}}&\underline{\textbf{0.919}}&\underline{\textbf{32.43}}&\underline{\textbf{0.945}}&\underline{\textbf{31.14}}&\underline{\textbf{0.937}}\\
      &DBSN&30.15&0.918&31.25&0.931&29.74&0.916\\
      &Noise2Void&27.82&0.870&28.65&0.885&27.12&0.858\\
      \cmidrule{2-8}
      
      &U-Net+N2V&29.15&0.899&30.33&0.918&28.55&0.893\\
      &Ours+N2V&\underline{29.57}&\underline{0.905}&\underline{30.81}&\underline{0.922}&\underline{29.21}&\underline{0.903}\\
      \bottomrule
      &Baseline, N2N&31.12&0.913&33.12&0.943&31.45&0.934\\
      \cmidrule{2-8}
      \multirow{4}{*}{Gaussian}&Laine19-mu&29.46&0.897&30.86&0.919&28.59&0.891\\
      \multirow{4}{*}{$\sigma \in [5,50]$}&Laine19-pme&\underline{\textbf{31.08}}&\underline{\textbf{0.911}}&\underline{\textbf{33.05}}&\underline{\textbf{0.940}}&\underline{\textbf{31.35}}&\underline{\textbf{0.931}}\\
      &DBSN&29.60&0.895&30.91&0.891&28.79&0.879\\
      &Noise2Void&27.77&0.873&28.76&0.881&26.95&0.849\\
      \cmidrule{2-8}
      
      &U-Net+N2V&29.22&0.898&30.56&0.917&28.44&0.889\\
      &Ours+N2V&\underline{29.78}&\underline{0.908}&\underline{31.22}&\underline{0.923}&\underline{29.21}&\underline{0.901}\\
      \bottomrule
    \end{tabular}}
    \caption{Quantitative comparison (PSNR/SSIM) results of dataset-based learning methods for Gaussian noise.}
    \label{tab:denoising_dataset}
  \end{center}
\end{table}
\subsection{Implementation Details}
\subsubsection{Training Details}
\textbf{1) Single-image based:} For Self2Self setting, the dropout rate of all dropout layers for Bernoulli sampling and regularization in the convolution is set to 0.7. The adam optimizer is used for training with the learning rate initialized to 0.0001 and 150,000 training steps. During testing, we run inference of each image 100 times and average them to get denoising results. 
\textbf{2) Dataset-based:} For training, we select images whose size is between 256$\times$256 and 512$\times$512 from training dataset, and then randomly crop 256$\times$256 patches as input. We use a batch size of 4 and adam optimizer with an initial learning rate of 0.0003 that is adjusted every iteration~\cite{laine2019high}. The number of iterations is 500,000. For denoising training strategy, we follow Noise2Void setting without 64$\times$64 patch extraction~\cite{krull2020probabilistic}. All our experiments are conducted on one NVIDIA Tesla V100 GPU.

\subsubsection{Dataset Details}
\textbf{1) Synthetic Noisy Datasets:} We consider two synthetic noise distributions, i.e. Gaussian noise with a fixed level $\sigma =25$ and Gaussian noise with varied noise levels $\sigma \in [5, 50]$. For single-image based evaluation, three datasets are used for performance evaluation, including Set14 (14 images)~\cite{zeyde2010single}, Kodak (24 images)~\cite{franzen1999kodak}, and McMaster (18 images)~\cite{zhang2011color}. For dataset-based, we adopt 50k images from Imagenet validation set~\cite{deng2009imagenet} as training dataset. The testsets are Set14 (14 images), Kodak (24 images), and BSD300 test set (100 images)~\cite{martin2001database}. \textbf{2) Real-world Noisy Datasets:} The denoising evaluation on realistic noisy images is conducted on the PolyU dataset~\cite{xu2018real}, containing 100 noisy-clean color image pairs. 70 image pairs are randomly selected for training (if method is dataset-based), while the remaining images are for testing.

\begin{figure*}[thpb]
\setlength\tabcolsep{0.6 pt}
\renewcommand{\arraystretch}{0.5}
\begin{center}
\subfloat[\normalsize{Gaussian $\sigma = 25$}]{
\resizebox{0.8\textwidth}{!}{
\tiny
\begin{tabular}{ccccccc}
\includegraphics[height=2cm,width=2cm]{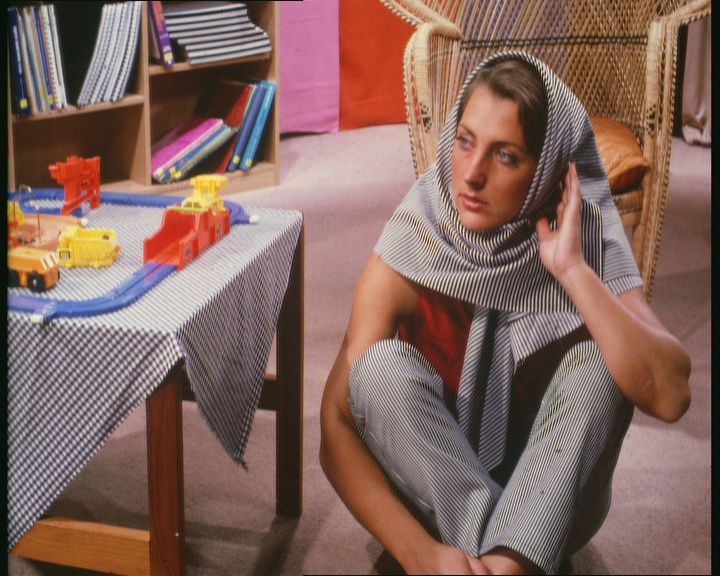}&\includegraphics[height=2cm,width=2cm]{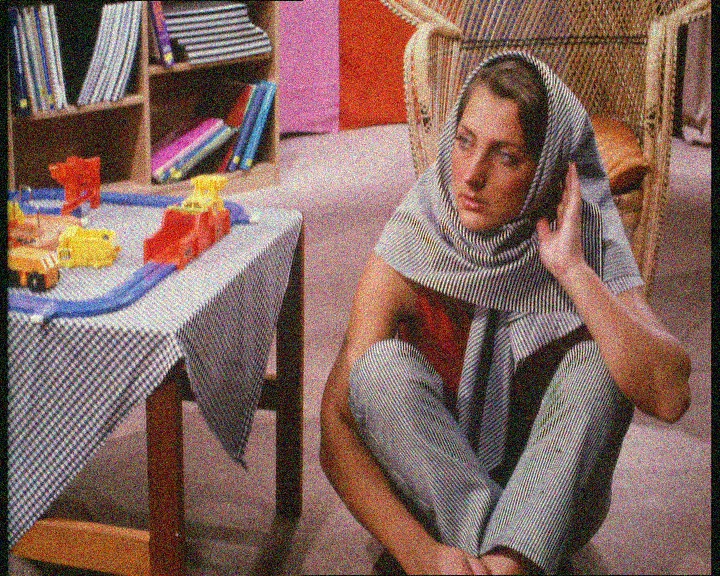}&\includegraphics[height=2cm,width=2cm]{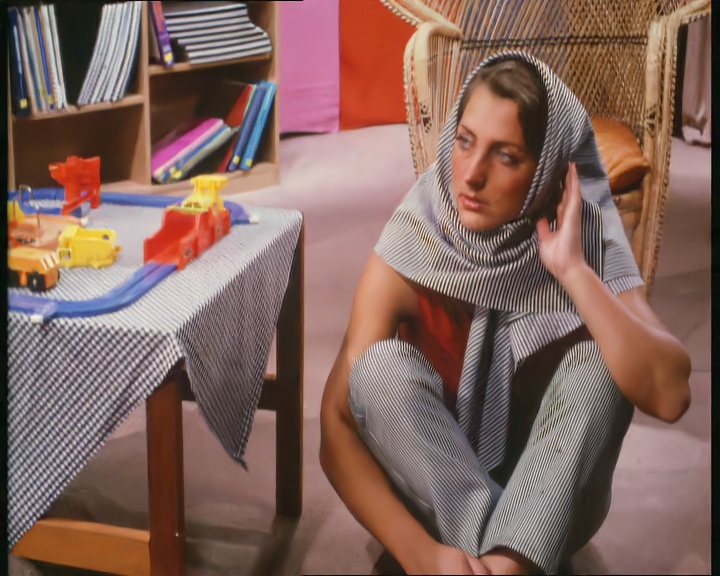}&\includegraphics[height=2cm,width=2cm]{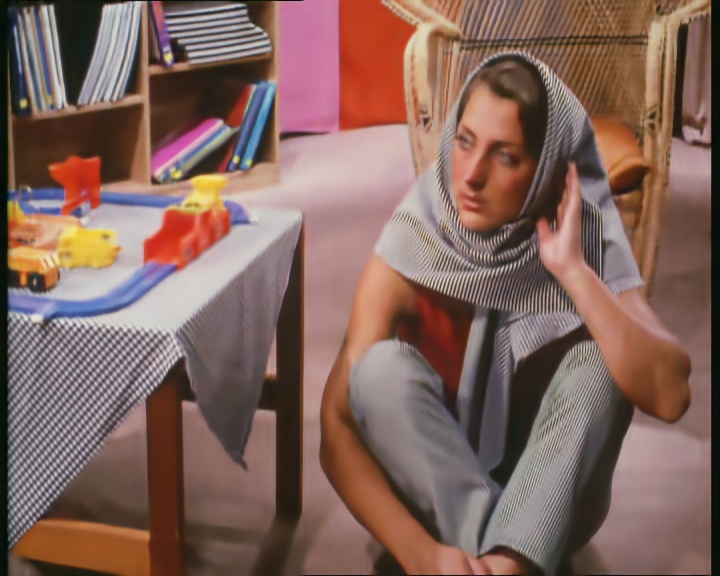}&\includegraphics[height=2cm,width=2cm]{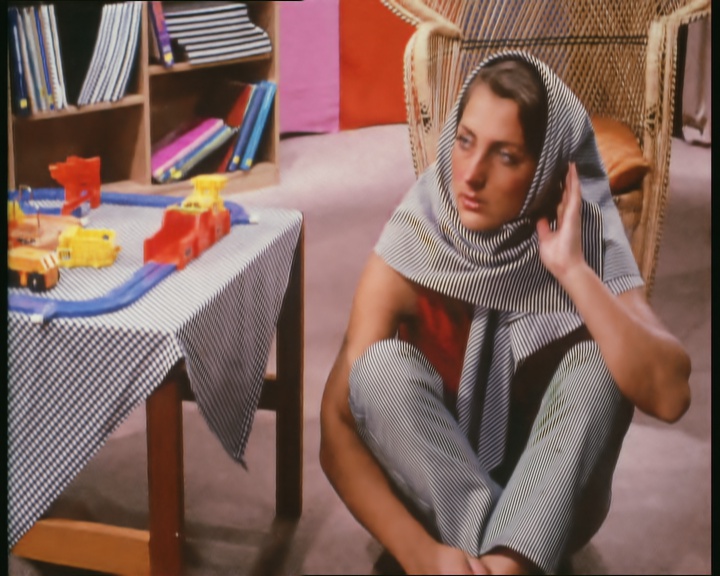}&\includegraphics[height=2cm,width=2cm]{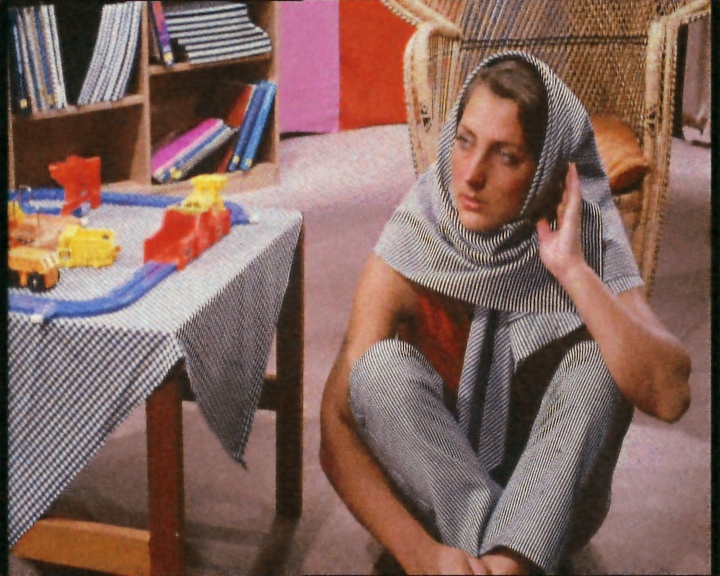}&\includegraphics[height=2cm,width=2cm]{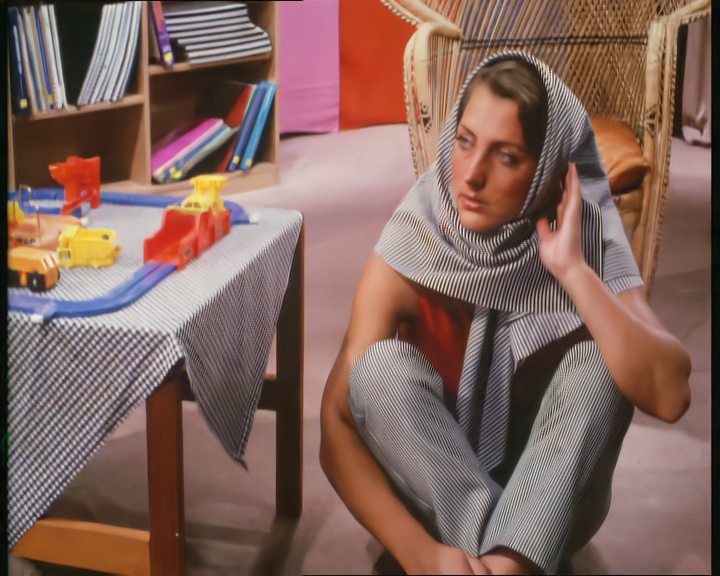}\\
\multirow{2}{*}{Set14-barbara}&Noisy&Noise2Noise&DIP&Self2Self&Noise2Void(1)&Laine19-mu\\
&20.37/0.644&31.47/0.957&27.44/0.875&32.02/0.96&28.59/0.915&30.61/0.949\\
\includegraphics[height=2cm,width=2cm]{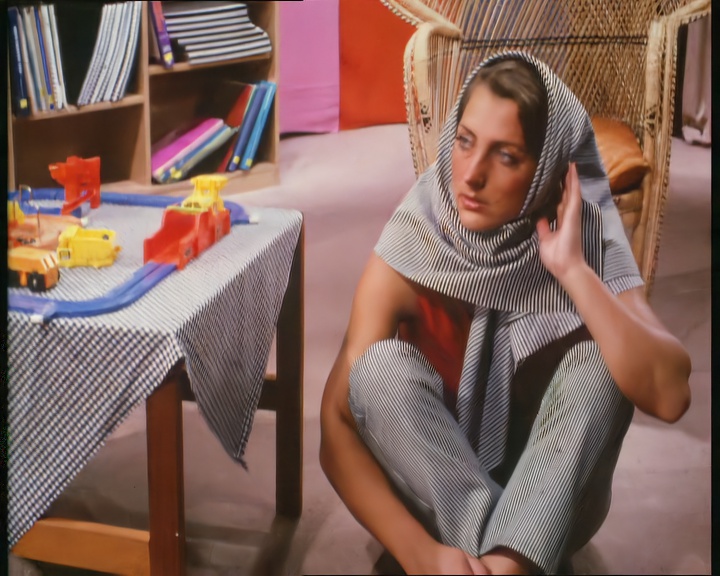}&\includegraphics[height=2cm,width=2cm]{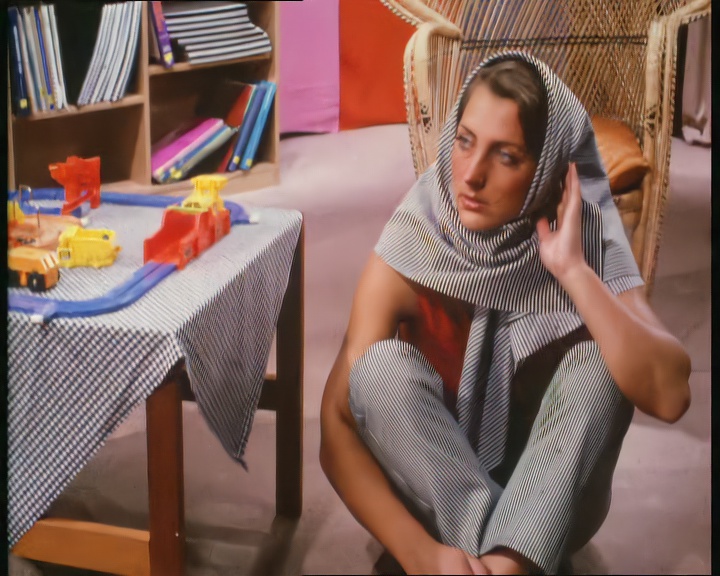}&\includegraphics[height=2cm,width=2cm]{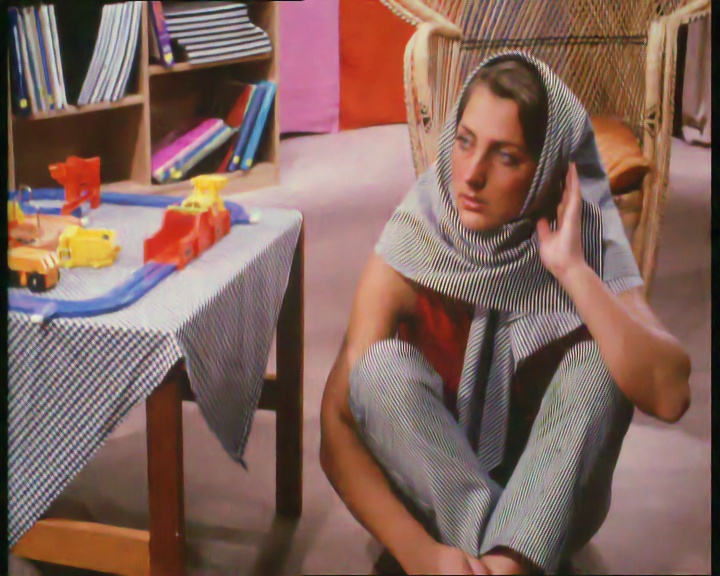}
&\includegraphics[height=2cm,width=2cm]{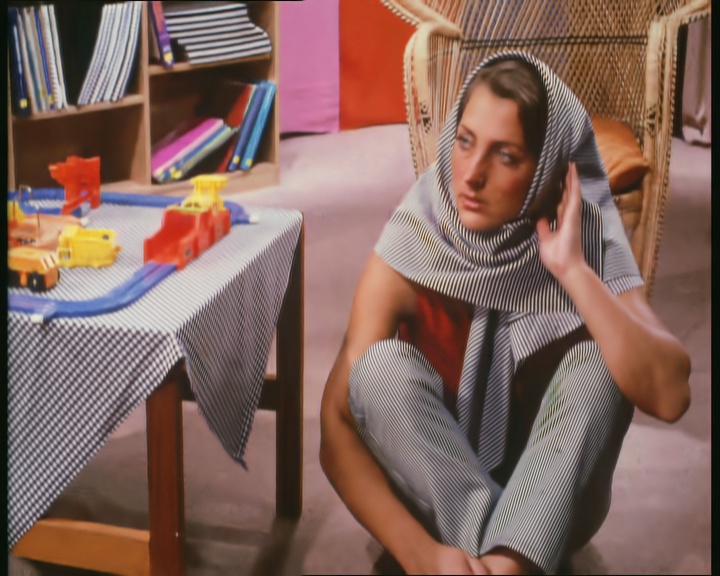}
&\includegraphics[height=2cm,width=2cm]{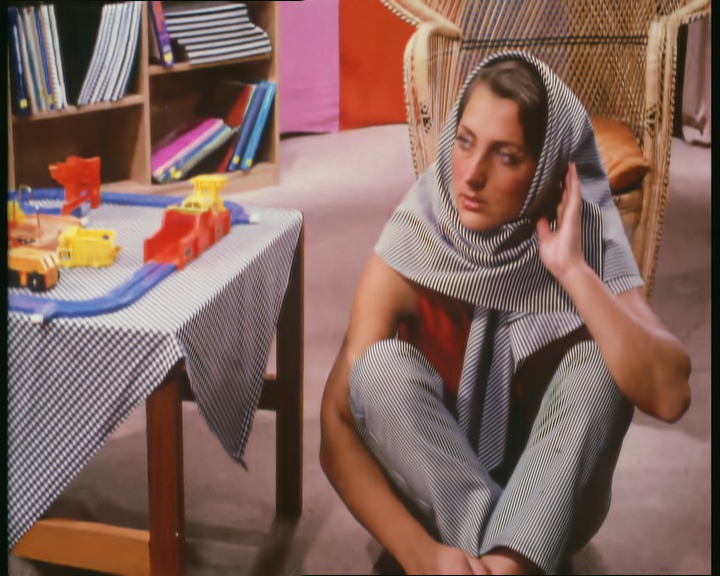}
&\includegraphics[height=2cm,width=2cm]{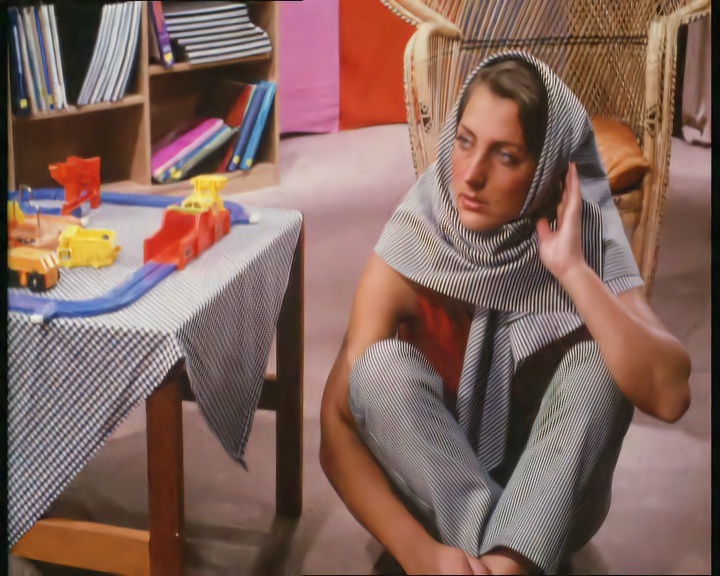}
&\includegraphics[height=2cm,width=2cm]{./imgs/gaussian_25/self_test_Set14_gaussian_sigma_25_denoising_my_unet_hybrid_self2self_dual_decomp_new_step_150000_v2/barbara/model_image/Self2Self_barbara_best.jpg}\\

Laine19-pme&DBSN&Noise2Void&U-Net+S2S(1)&Ours+S2S(1)&U-Net+N2V&Ours+N2V\\
31.56/\underline{0.957}&30.96/0.952&27.53/0.894&32.2/0.962&\underline{\textbf{32.36}}/\underline{\textbf{0.963}}&30.28/0.943&\underline{32.2}/0.945\\
\end{tabular}}}

\subfloat[\normalsize{Gaussian $\sigma \in [5,50]$}]{
\resizebox{0.8\textwidth}{!}{
\huge
\begin{tabular}{ccccccc}
\includegraphics[scale=0.2]{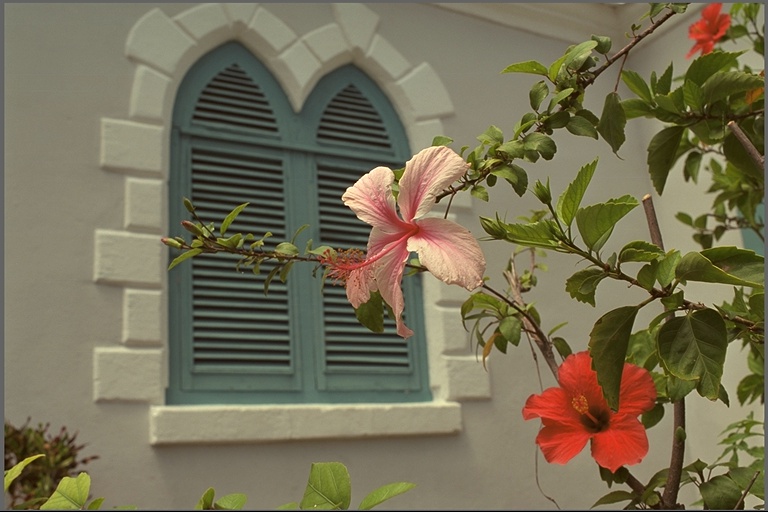}&\includegraphics[scale=0.2]{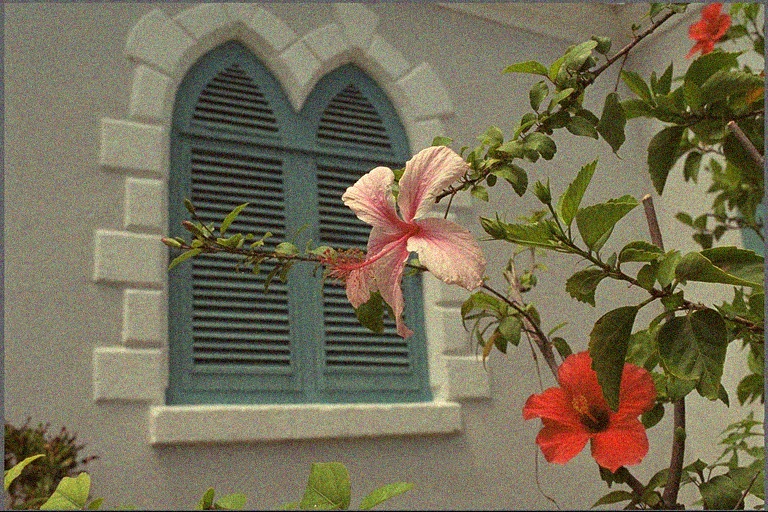}&\includegraphics[scale=0.2]{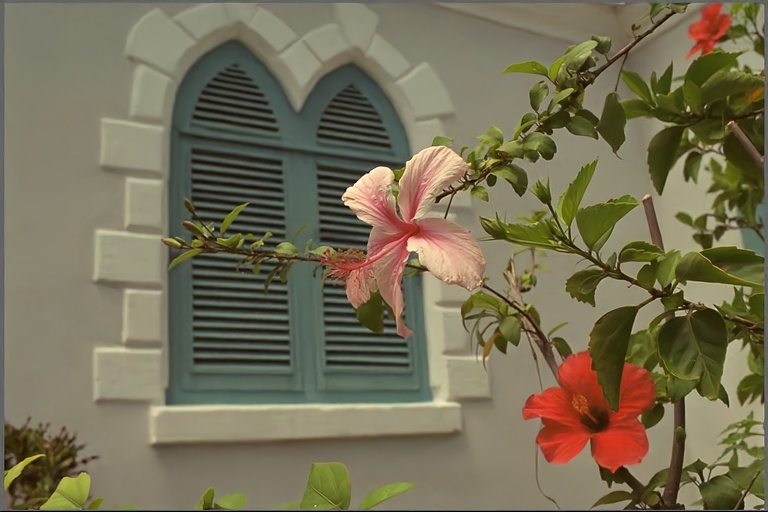}&\includegraphics[scale=0.2]{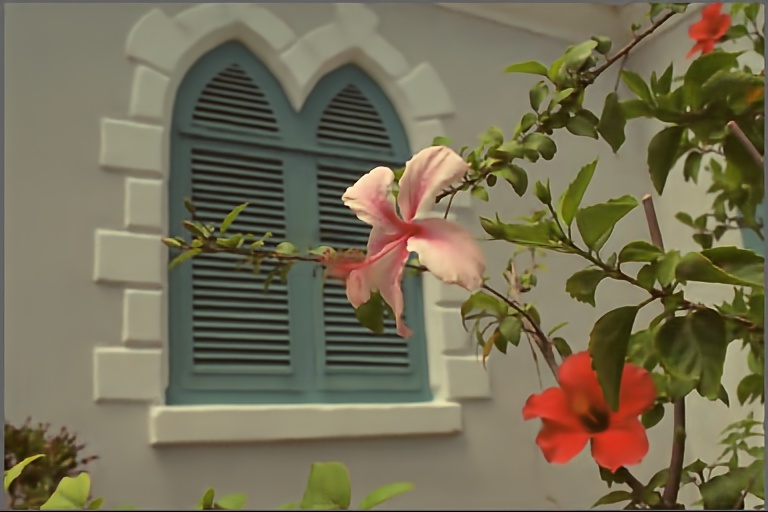}&\includegraphics[scale=0.2]{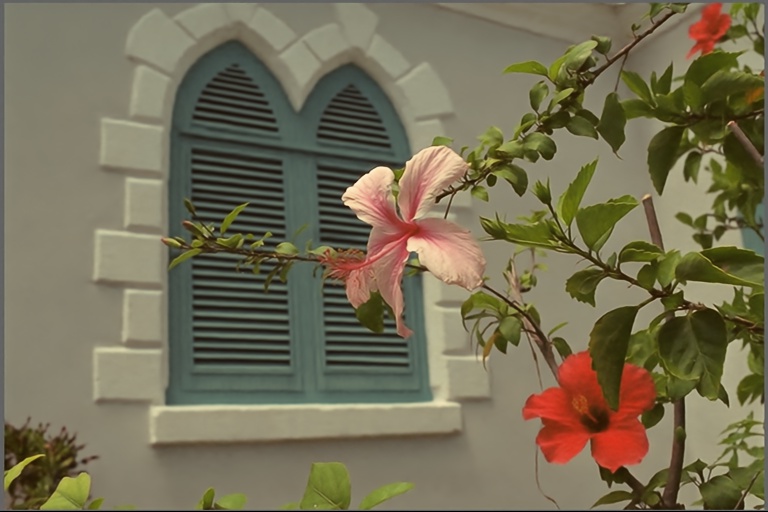}&\includegraphics[scale=0.2]{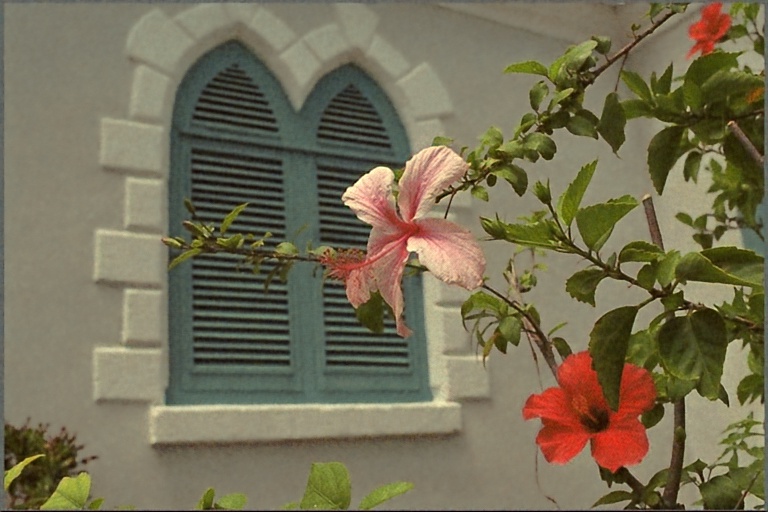}&\includegraphics[scale=0.2]{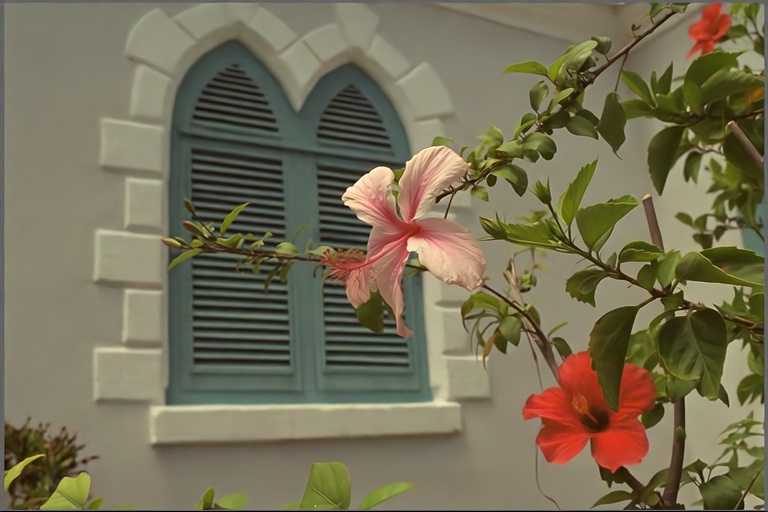}\\
\multirow{2}{*}{koda-kodim07}&Noisy&Noise2Noise&DIP&Self2Self&Noise2Void(1)&Laine19-mu\\
&24.36/0.686&36.81/0.986&33.24/0.975&36.04/\underline{\textbf{0.986}}&33.04/0.968&35.17/0.982\\
\includegraphics[scale=0.2]{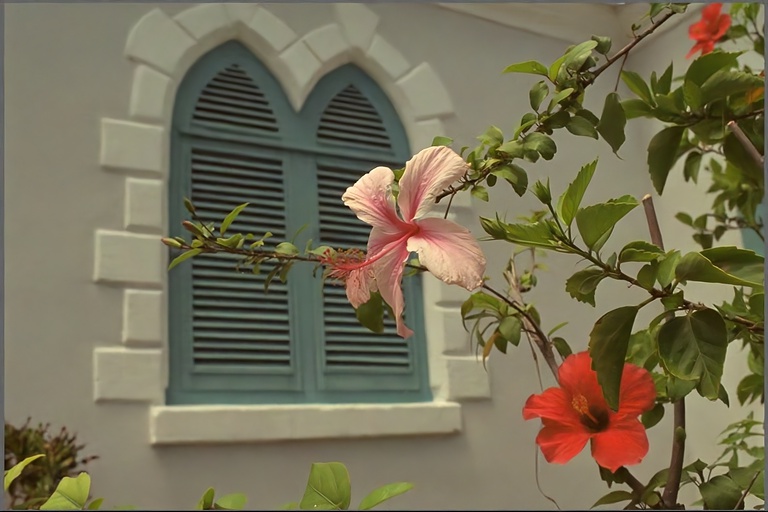}&\includegraphics[scale=0.2]{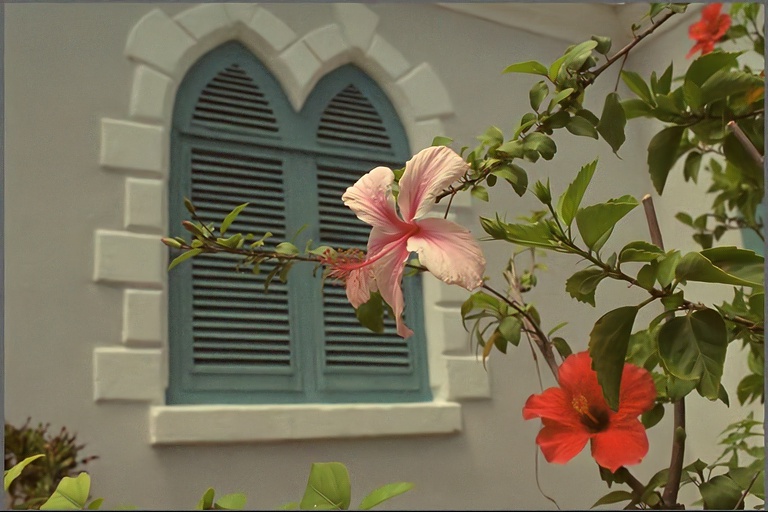}&\includegraphics[scale=0.2]{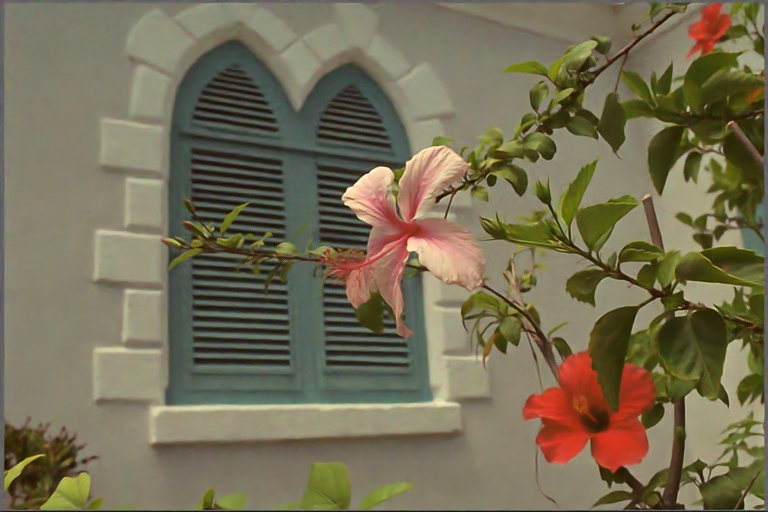}

&\includegraphics[scale=0.2]{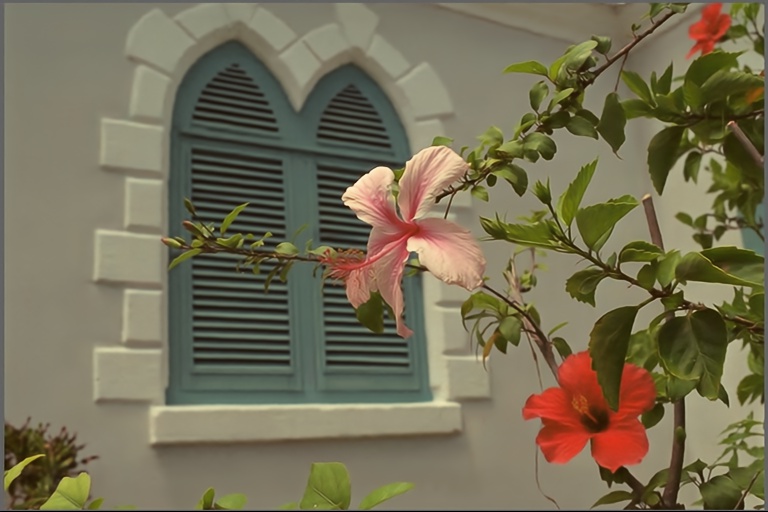}

&\includegraphics[scale=0.2]{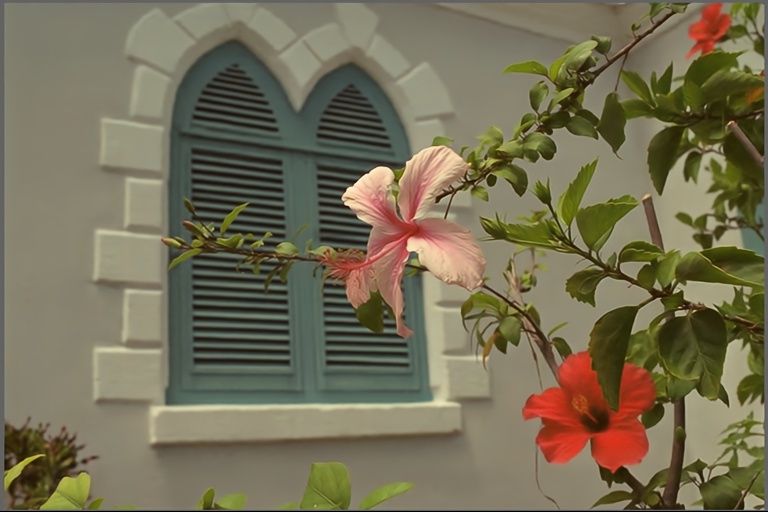}

&\includegraphics[scale=0.2]{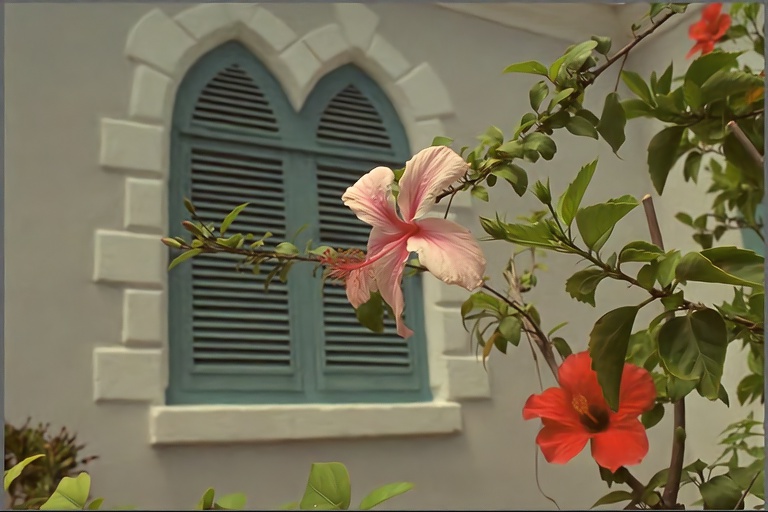}

&\includegraphics[scale=0.2]{./imgs/gaussian_5_50/self_test_koda_gaussian_sigma_range_5_50_denoising_my_unet_hybrid_self2self_dual_decomp_new_step_150000_v2/kodim07/model_image/Self2Self_kodim07_best.jpg}\\

Laine19-pme&DBSN&Noise2Void&U-Net+S2S(1)&Ours+S2S(1)&U-Net+N2V&Ours+N2V\\
\underline{\textbf{36.77}}/\underline{0.985}&34.87/0.973&32.27/0.964
&36.16/0.986
&\underline{36.28}/\underline{\textbf{0.986}}
&34.67/0.98
&36.16/0.981\\
\end{tabular}}}
\caption{Comparison of denoising results in the setting of Gaussian $\sigma=25$ and $\sigma \in [5,50]$.} 
\label{fig:syn_noise_data} 
\end{center}
\end{figure*}

\begin{table*}[ht!]
  \begin{center}
    \resizebox{\textwidth}{!}{%
    \begin{tabular}{cccccccccc}
    \toprule 
      \multirow{2}{*}{\textbf{Evaluation metrics}} & \multicolumn{5}{c}{\textbf{Single-image based learning methods}}& & \multicolumn{3}{c}{\textbf{Dataset-based learning methods}}\\
      \cmidrule(lr){2-6} \cmidrule(lr){7-10}
      & DIP & Noise2Void(1)&Self2Self&U-Net+S2S(1)&Ours+S2S(1)&&Noise2Void&U-Net+N2V&Ours+N2V\\
      \toprule
      PSNR$\uparrow$& 37.35& 35.14&37.95&37.17&\underline{\textbf{38.10}}& &35.46& 35.47&\underline{35.99}\\
      SSIM$\uparrow$& 0.982& 0.958&\underline{\textbf{0.984}}&0.954&\underline{\textbf{0.984}}& &0.958& 0.958&\underline{0.966}\\
      \bottomrule 
    \end{tabular}}
    \caption{Quantitative comparsion on PolyU dataset}
    \label{tab:polyu}
  \end{center}
\end{table*}

\subsection{Experimental Results}
\subsubsection{Comparison on Synthetic Noisy Images}
We use Noise2Noise (N2N)~\cite{lehtinen2018noise2noise} as a baseline benchmark, reproduced by officially-released pre-trained model. This section is separated into two different comparison parts as follows:

\textbf{1) Comparison to single-image-based learning methods:} We compare our approach with popular single-image based learning methods, i.e. DIP~\cite{ulyanov2018deep}, Self2Self~\cite{quan2020self2self}, and single-image version of Noise2Void~\cite{krull2019noise2void}, denoted by Noise2Void(1). Recall that Noise2Void is trained on unorganized noisy images, Noise2Noise on paired noisy images, and the rest on single noisy image. We reproduce results of compared approaches on our synthetic noise testsets by utilizing their official implementations. Our network structure cooperated with Self2Self training scheme is described by Ours+S2S(1). We also conduct ablation study by replacing MGRConv with vanilla convolution in the encoder, which is regarded as U-Net+S2S(1). \textbf{Noted that the training parameters between ours and Self2Self reproduction are the same.}
From Table~\ref{tab:denoising_single}, our approach outperforms DIP and Noise2Void(1) by a large margin and even sometimes it has better performance than Noise2Noise baseline. Surprisingly, Noise2Noise does not perform well on McMaster dataset, which seems to be a common scenario among dataset-based approachs trained by ImageNet validation set. Compared to Self2Self and U-Net+S2S(1), the MGRConv based architecture introduces stable denoising performance facilitation, which shows the effectiveness of our proposed module. See Fig.~\ref{fig:syn_noise_data} for visual comparison. In Fig.~\ref{fig:syn_noise_data} (a), our MGRConv can boost denoising performance significantly by 0.34 in PSNR compared to Self2Self and even performs better than Noise2Noise, and dataset-based state-of-the-art approaches, including Laine et al. and DBSN.

\textbf{2) Comparison to dataset-based learning methods:} Our method is also compared to dataset-based learning methods, including Noise2Void~\cite{krull2019noise2void}, DBSN~\cite{wu2020unpaired} and Laine19~\cite{laine2019high}. We use the Laine19 pre-trained model provided by authors, in which result without post-processing is denoted as Laine19-mu, while post-processed posterior mean estimation result is described by Laine19-pme. For Noise2Void and DBSN, we reproduce results using official implementations, where Noise2Void is trained with BSD300 training set, and DBSN with ImageNet validation set. Similarly, our approach and its ablation study are Ours+N2V and U-Net+N2V respectively, which are trained on noisy images instead of unorganized noisy images.
\begin{figure}[thpb!]
\Large
\setlength\tabcolsep{0.6 pt}
\renewcommand{\arraystretch}{0.5}
\begin{center}
\resizebox{\columnwidth}{!}{
\begin{tabular}{cccc}
\includegraphics[scale=0.2]{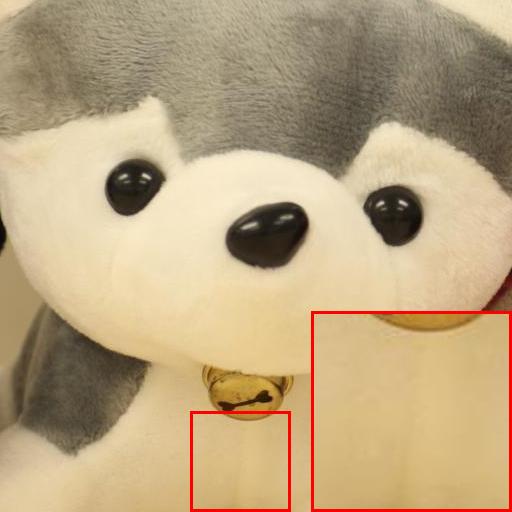}&\includegraphics[scale=0.2]{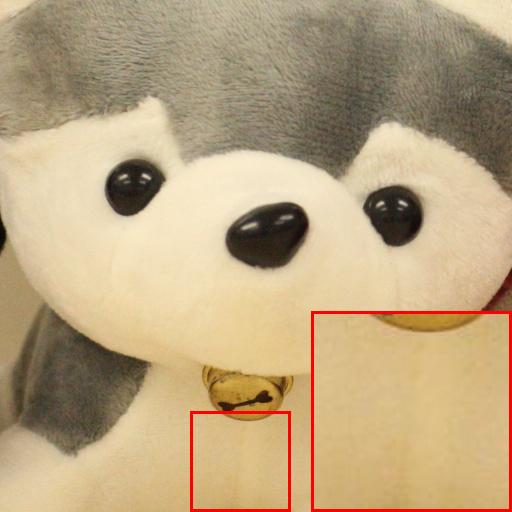}&\includegraphics[scale=0.2]{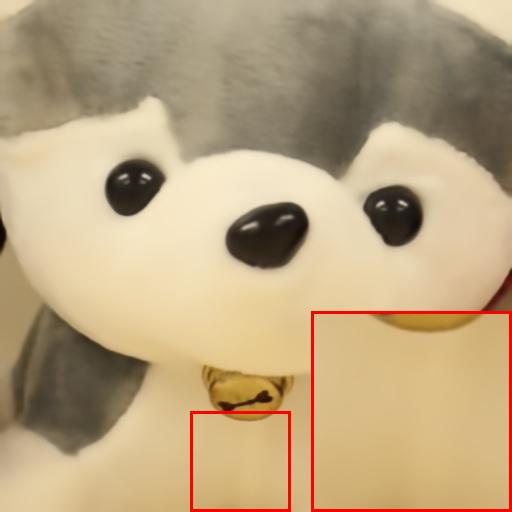}&\includegraphics[scale=0.2]{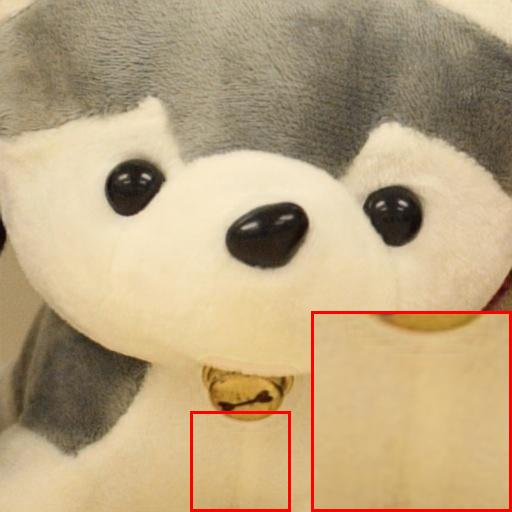}\\
\multirow{2}{*}{GT}&Noisy&DIP&Noise2Void(1)\\
&37.0/0.984&34.85/0.971&36.35/\underline{0.983}\\
\includegraphics[scale=0.2]{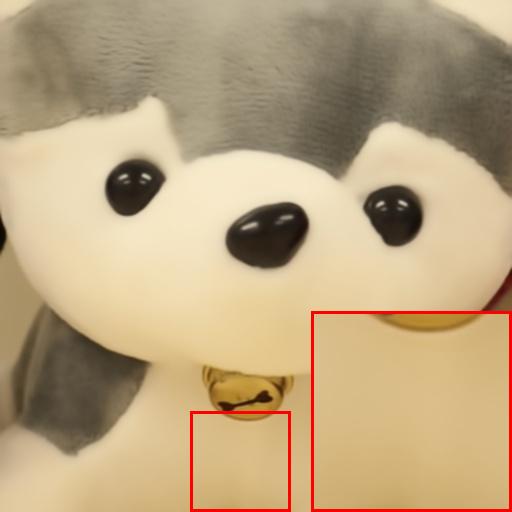}&\includegraphics[scale=0.2]{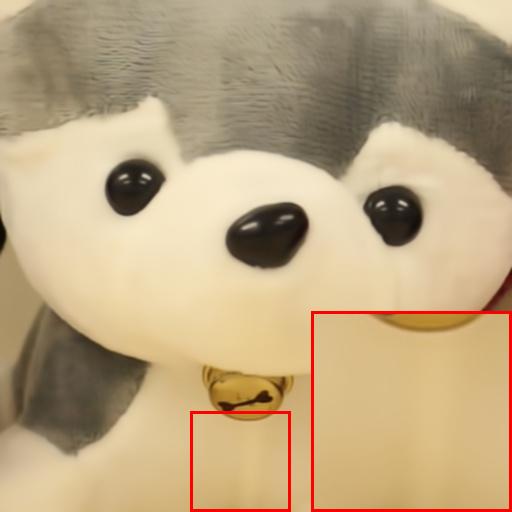}&\includegraphics[scale=0.2]{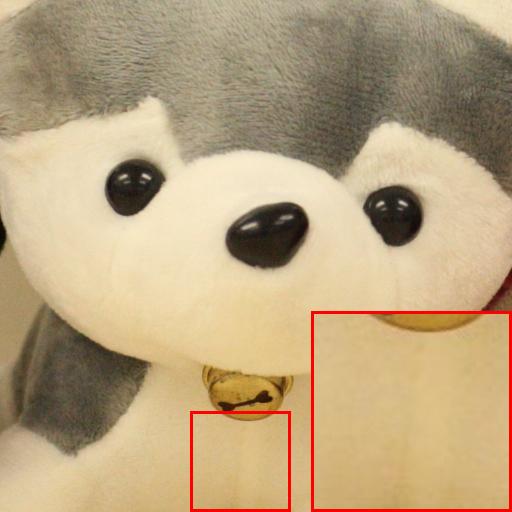}&\includegraphics[scale=0.2]{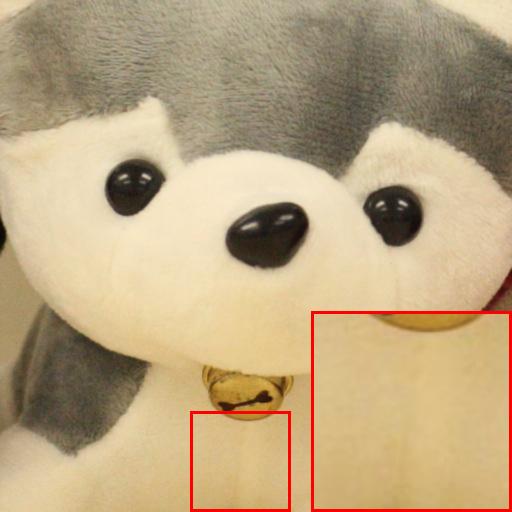}\\
Self2Self&Ours+S2S(1)&Noise2Void&Ours+N2V\\
35.81/0.979&\underline{36.73}/0.982&36.68/0.984&\underline{\textbf{37.33}}/\underline{\textbf{0.986}}\\
\end{tabular}}
\caption{Visual comparison of denoising results on PolyU dataset.} 
\label{fig:polyu_data} 
\end{center}
\end{figure}
\begin{table}[th]
  \begin{center}
    \resizebox{0.7\columnwidth}{!}{%
    \begin{tabular}{ccccc}
      \toprule 
      \multirow{2}{*}{Methods}&\multirow{2}{*}{Metrics}&\multicolumn{3}{c}{Dropping Ratio}\\
      \cmidrule{3-5} 
      &&50\%&70\%&90\%\\
      \midrule
      \multirow{2}{*}{PConv}&PSNR$\uparrow$&34.01&30.64 &\textbf{25.19}\\
                            &SSIM$\uparrow$&0.9499&0.9083 &\textbf{0.7864}\\
                            \midrule
      \multirow{2}{*}{LBAM}&PSNR$\uparrow$&34.01&30.55&24.82\\
                           &SSIM$\uparrow$&\textbf{0.9510}&\underline{0.9090}&0.7762\\
                           \midrule
      \multirow{2}{*}{GatedConv}&PSNR$\uparrow$&\underline{34.07}&\underline{30.68}&25.04\\
                                &SSIM$\uparrow$&0.9506&\underline{0.9090}&0.7827\\
                                \midrule
      \multirow{2}{*}{Ours}&PSNR$\uparrow$&\textbf{34.09}&\textbf{30.70}&\underline{25.07}\\
                           &SSIM$\uparrow$&\underline{0.9507}&\textbf{0.9091}&\underline{0.7855}\\
      \bottomrule 
    \end{tabular}}
    \caption{PSNR/SSIM of inpainting results of our network architecture with Self2Self setting on Set12. For each dropping ratio, the best results are marked in bold and the second ones are underlined.}
    \label{tab:inpainting}
  \end{center}
\end{table}
See Table~\ref{tab:denoising_dataset} and Fig.~\ref{fig:syn_noise_data} for comparison. Our network architecture with plain N2V training strategy suppresses Noise2Void approach greatly. The ablation study of Ours+N2V and U-Net+N2V also demonstrates the superiority of our MGRConv, which can have comprehensive application in self-denoising.

\subsubsection{Comparison on Real-world Noisy Images}
Following aforementioned presentation style, our denoising framework is compared with DIP, Noise2Void(1), Self2Self, and Noise2Void. We reproduce compared methods by published codes, in which dataset-based methods are trained by 70 image pairs randomly picked up from PolyU dataset. In Table~\ref{tab:polyu}, our method obviously outperforms both single-image based and dataset-based learning approaches. Since the MGRConv can make network learn dynamic mask and treat each pixel in each channel unequally, it fits the real-world situation of uneven noise distribution and provides specific adaption learning for given noisy image. In Fig.~\ref{fig:polyu_data}, our method not only preserves furry details, but also removes stain from the fur.

\subsubsection{Comparison on Inpainting Task}

\begin{figure}[thpb]
\tiny
\setlength\tabcolsep{0.6 pt}
\renewcommand{\arraystretch}{0.5}
\begin{center}
\resizebox{\columnwidth}{!}{
\begin{tabular}{ccc}
\includegraphics[height=2cm,width=2cm]{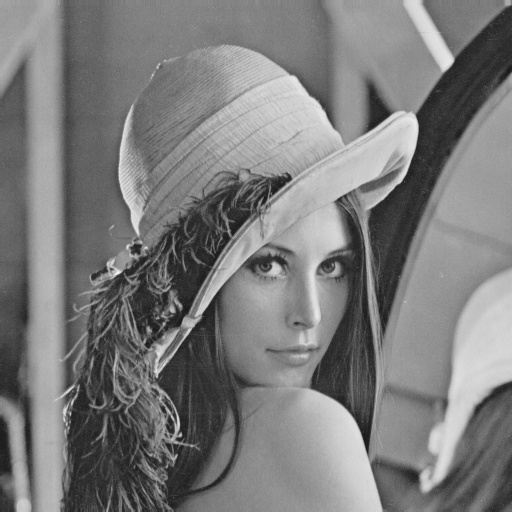}&\includegraphics[height=2cm,width=2cm]{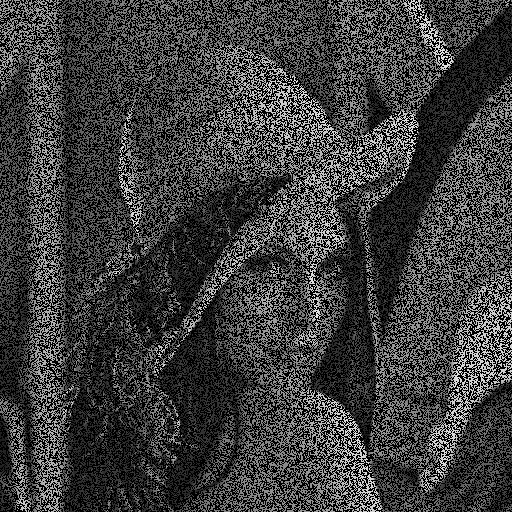}&\includegraphics[height=2cm,width=2cm]{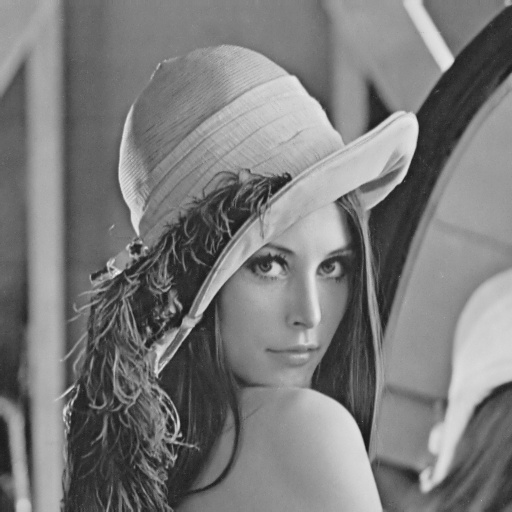}\\
\multirow{2}{*}{Set12-08}&Input(50.0\% dropped)&PConv\\
&8.69/0.0599&37.73/0.9559\\
\includegraphics[height=2cm,width=2cm]{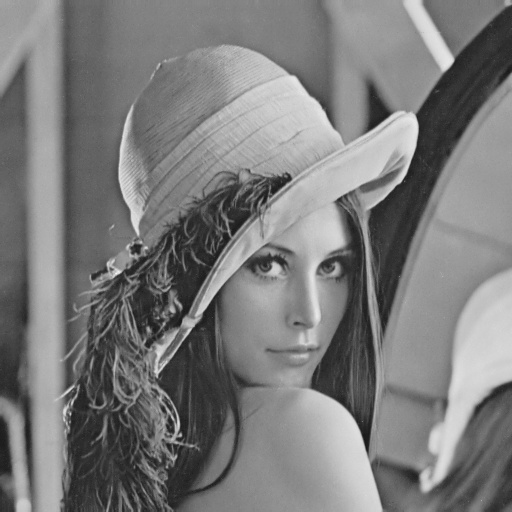}&\includegraphics[height=2cm,width=2cm]{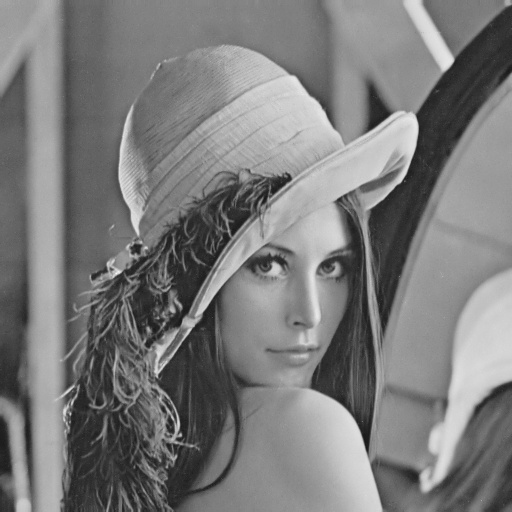}&\includegraphics[height=2cm,width=2cm]{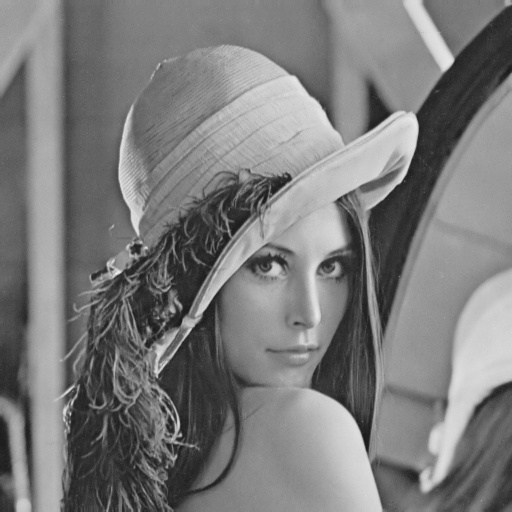}\\
LBAM&GatedConv&Ours\\
\underline{37.83}/\textbf{0.9569}&37.77/0.956&\textbf{37.84}/\underline{0.9564}\\
\end{tabular}}
\caption{Example inpainting cases of qualitative comparison.} 
\label{fig:inpainting_data_s2s} 
\end{center}
\end{figure}

\begin{figure}[h]
\tiny
\setlength\tabcolsep{0.6 pt}
\renewcommand{\arraystretch}{0.5}
\begin{center}
\resizebox{\columnwidth}{!}{
\begin{tabular}{ccc}
\includegraphics[height=2cm,width=2cm]{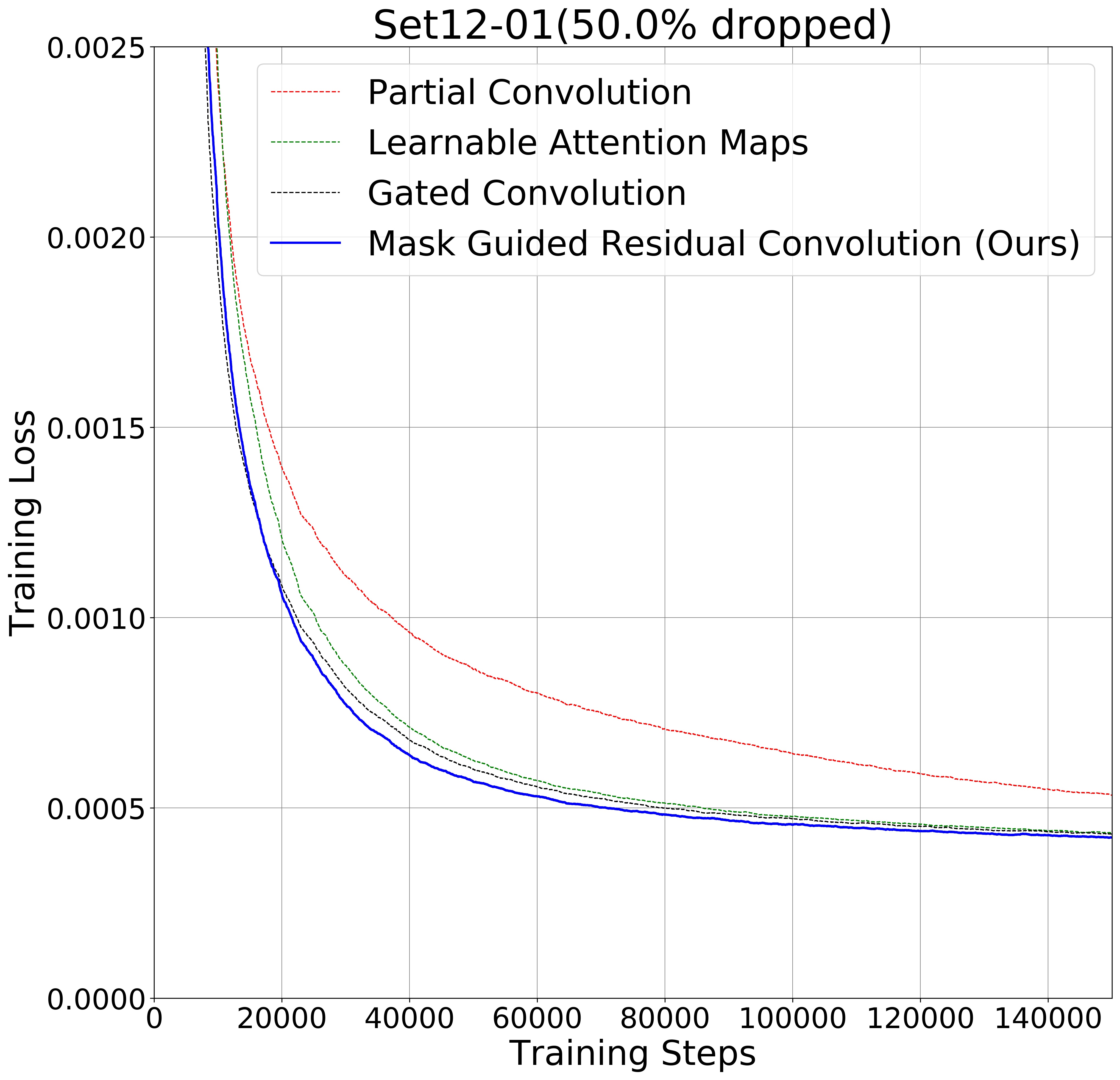}&\includegraphics[height=2cm,width=2cm]{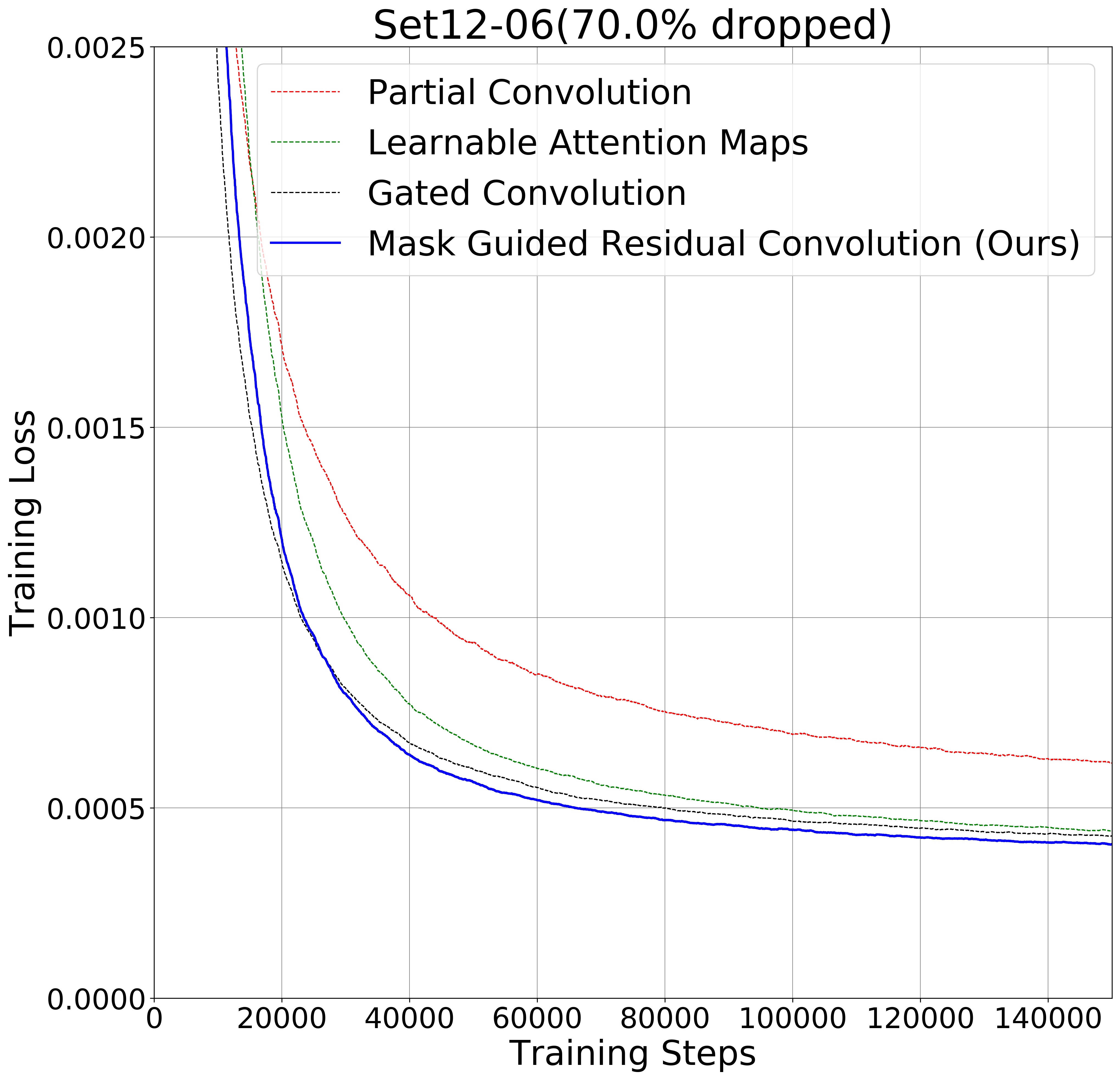}&\includegraphics[height=2cm,width=2cm]{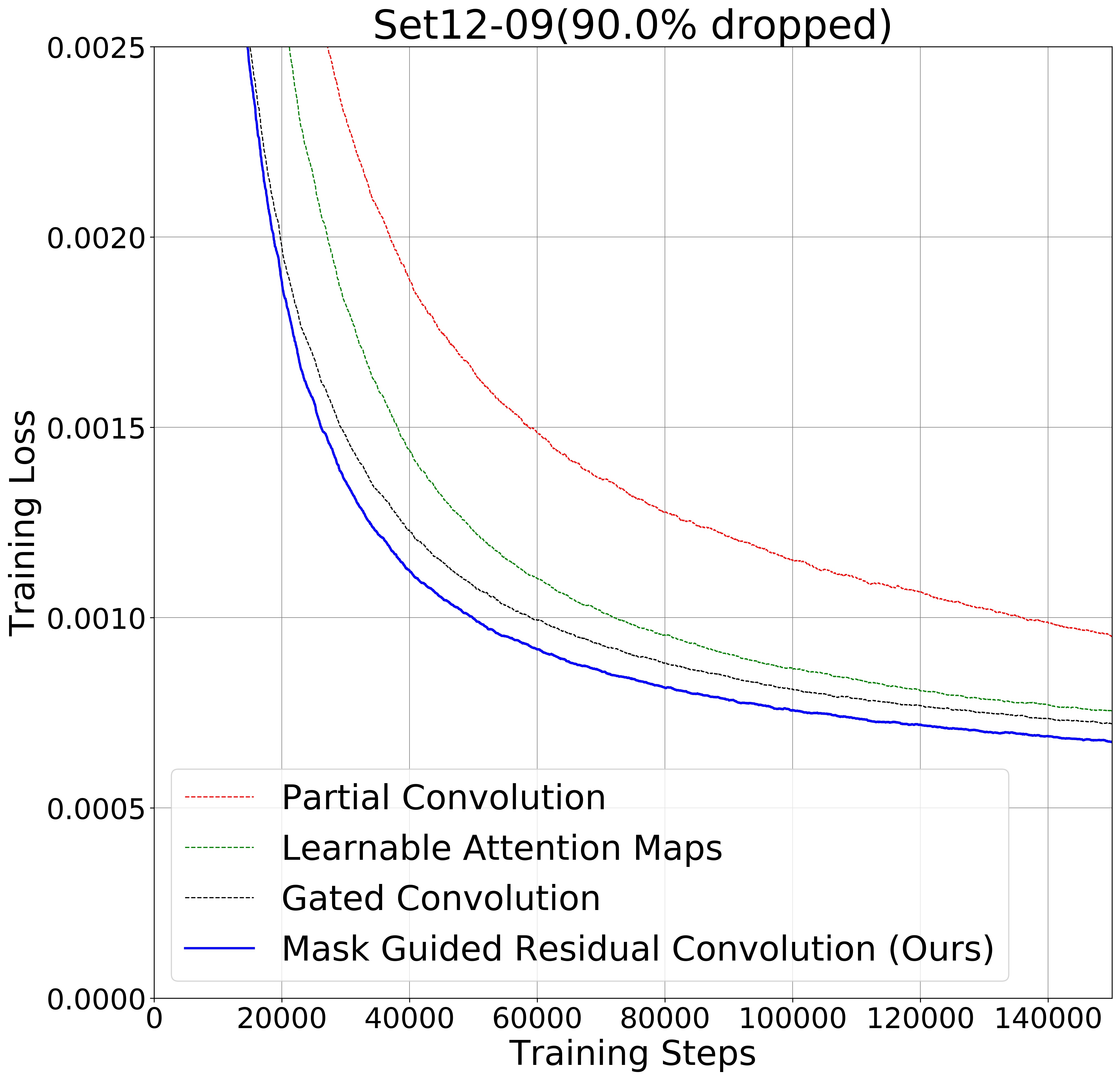}\\
\end{tabular}}
\caption{Comparison of training process between MGRConv and other inpainting convolutions on different images.}
\label{fig:training_loss_compare} 
\end{center}
\end{figure}

To further compare our MGRConv with other inpainting convolutions, we generate corrupted images by randomly dropping pixels with ratios 50\%, 70\%, and 90\% respectively to conduct inpainting experiments on Set12 dataset using Self2Self setting. We replace MGRConv in our network with PConv, LBAM, and GatedConv, in which we use forward layer of LBAM. See Table~\ref{tab:inpainting} and Fig.~\ref{fig:inpainting_data_s2s} for quantitative and visual comparison. We also visualize training process to illustrate the superiority of our proposed network module. The Fig.~\ref{fig:training_loss_compare} shows the training curves of MGRConv and other inpainting convolutions. Clearly, our MGRConv converges much faster than PConv does. The final status of training convergence and inpainting results, shown in Fig.~\ref{fig:training_loss_compare} and Table~\ref{tab:inpainting}, indicate that MGRConv performs much better than others. Furthermore, the quantitative comparison of GFLOPs of different inpainting convolutions is disclosed as: PConv (0.368), LBAM (0.384), GatedConv (0.345), MGRConv (0.355). Noted that our MGRConv obtains the most excellent computation cost compared to PConv and LBAM, and is on par with the nonstrictly image/mask separated convolution, i.e. GatedConv.

\section{Conclusion}
This paper proposes Mask Guided Residual Convolution (MGRConv) for blind-spot based self-denoising. By introducing MGRConv and modeling blind-spot masking strategy as inpainting procedure, upgraded neural network can be more effective and achieve better performance. Conducted experiments show that our MGRConv can be a helpful plug-and-play assistance for comprehensive blind-spot based denoising variances in generating satisfactory denoising results.

\section{Acknowledgments}
Thank Qi Song, Zheng Wang, and Junjie Hu for comments or discussions. Thank Yuejin Li for his support in using GPU clusters.

\bibliography{Ref}

\begin{thebibliography}{35}
\providecommand{\natexlab}[1]{#1}

\bibitem[{Batson and Royer(2019)}]{batson2019noise2self}
Batson, J.; and Royer, L. 2019.
\newblock Noise2self: Blind denoising by self-supervision.
\newblock In \emph{International Conference on Machine Learning}, 524--533.
  PMLR.

\bibitem[{Buades, Coll, and Morel(2005)}]{buades2005non}
Buades, A.; Coll, B.; and Morel, J.-M. 2005.
\newblock A non-local algorithm for image denoising.
\newblock In \emph{2005 IEEE Computer Society Conference on Computer Vision and
  Pattern Recognition (CVPR'05)}, volume~2, 60--65. IEEE.

\bibitem[{Chen et~al.(2021)Chen, Wang, Guo, Xu, Deng, Liu, Ma, Xu, Xu, and
  Gao}]{chen2021pre}
Chen, H.; Wang, Y.; Guo, T.; Xu, C.; Deng, Y.; Liu, Z.; Ma, S.; Xu, C.; Xu, C.;
  and Gao, W. 2021.
\newblock Pre-trained image processing transformer.
\newblock In \emph{Proceedings of the IEEE/CVF Conference on Computer Vision
  and Pattern Recognition}, 12299--12310.

\bibitem[{Dabov et~al.(2007)Dabov, Foi, Katkovnik, and
  Egiazarian}]{dabov2007image}
Dabov, K.; Foi, A.; Katkovnik, V.; and Egiazarian, K. 2007.
\newblock Image denoising by sparse 3-D transform-domain collaborative
  filtering.
\newblock \emph{IEEE Transactions on image processing}, 16(8): 2080--2095.

\bibitem[{Deng et~al.(2009)Deng, Dong, Socher, Li, Li, and
  Fei-Fei}]{deng2009imagenet}
Deng, J.; Dong, W.; Socher, R.; Li, L.-J.; Li, K.; and Fei-Fei, L. 2009.
\newblock Imagenet: A large-scale hierarchical image database.
\newblock In \emph{2009 IEEE conference on computer vision and pattern
  recognition}, 248--255. Ieee.

\bibitem[{Franzen(1999)}]{franzen1999kodak}
Franzen, R. 1999.
\newblock Kodak lossless true color image suite.
\newblock \emph{source: http://r0k. us/graphics/kodak}, 4(2).

\bibitem[{Gu et~al.(2019)Gu, Li, Gool, and Timofte}]{gu2019self}
Gu, S.; Li, Y.; Gool, L.~V.; and Timofte, R. 2019.
\newblock Self-guided network for fast image denoising.
\newblock In \emph{Proceedings of the IEEE/CVF International Conference on
  Computer Vision}, 2511--2520.

\bibitem[{Gu et~al.(2014)Gu, Zhang, Zuo, and Feng}]{gu2014weighted}
Gu, S.; Zhang, L.; Zuo, W.; and Feng, X. 2014.
\newblock Weighted nuclear norm minimization with application to image
  denoising.
\newblock In \emph{Proceedings of the IEEE conference on computer vision and
  pattern recognition}, 2862--2869.

\bibitem[{Guo et~al.(2019)Guo, Yan, Zhang, Zuo, and Zhang}]{guo2019toward}
Guo, S.; Yan, Z.; Zhang, K.; Zuo, W.; and Zhang, L. 2019.
\newblock Toward convolutional blind denoising of real photographs.
\newblock In \emph{Proceedings of the IEEE/CVF Conference on Computer Vision
  and Pattern Recognition}, 1712--1722.

\bibitem[{Huang et~al.(2021)Huang, Li, Jia, Lu, and
  Liu}]{huang2021neighbor2neighbor}
Huang, T.; Li, S.; Jia, X.; Lu, H.; and Liu, J. 2021.
\newblock Neighbor2Neighbor: Self-Supervised Denoising from Single Noisy
  Images.
\newblock In \emph{Proceedings of the IEEE/CVF Conference on Computer Vision
  and Pattern Recognition}, 14781--14790.

\bibitem[{Krull, Buchholz, and Jug(2019)}]{krull2019noise2void}
Krull, A.; Buchholz, T.-O.; and Jug, F. 2019.
\newblock Noise2void-learning denoising from single noisy images.
\newblock In \emph{Proceedings of the IEEE/CVF Conference on Computer Vision
  and Pattern Recognition}, 2129--2137.

\bibitem[{Krull et~al.(2020)Krull, Vi{\v{c}}ar, Prakash, Lalit, and
  Jug}]{krull2020probabilistic}
Krull, A.; Vi{\v{c}}ar, T.; Prakash, M.; Lalit, M.; and Jug, F. 2020.
\newblock Probabilistic noise2void: Unsupervised content-aware denoising.
\newblock \emph{Frontiers in Computer Science}, 2: 5.

\bibitem[{Laine et~al.(2019)Laine, Karras, Lehtinen, and Aila}]{laine2019high}
Laine, S.; Karras, T.; Lehtinen, J.; and Aila, T. 2019.
\newblock High-quality self-supervised deep image denoising.
\newblock \emph{Advances in Neural Information Processing Systems}, 32:
  6970--6980.

\bibitem[{Lefkimmiatis(2018)}]{lefkimmiatis2018universal}
Lefkimmiatis, S. 2018.
\newblock Universal denoising networks: a novel CNN architecture for image
  denoising.
\newblock In \emph{Proceedings of the IEEE conference on computer vision and
  pattern recognition}, 3204--3213.

\bibitem[{Lehtinen et~al.(2018)Lehtinen, Munkberg, Hasselgren, Laine, Karras,
  Aittala, and Aila}]{lehtinen2018noise2noise}
Lehtinen, J.; Munkberg, J.; Hasselgren, J.; Laine, S.; Karras, T.; Aittala, M.;
  and Aila, T. 2018.
\newblock Noise2Noise: Learning Image Restoration without Clean Data.
\newblock In \emph{ICML}.

\bibitem[{Liu et~al.(2018)Liu, Reda, Shih, Wang, Tao, and
  Catanzaro}]{liu2018image}
Liu, G.; Reda, F.~A.; Shih, K.~J.; Wang, T.-C.; Tao, A.; and Catanzaro, B.
  2018.
\newblock Image inpainting for irregular holes using partial convolutions.
\newblock In \emph{Proceedings of the European Conference on Computer Vision
  (ECCV)}, 85--100.

\bibitem[{Mao, Shen, and Yang(2016)}]{mao2016image}
Mao, X.; Shen, C.; and Yang, Y.-B. 2016.
\newblock Image restoration using very deep convolutional encoder-decoder
  networks with symmetric skip connections.
\newblock \emph{Advances in neural information processing systems}, 29:
  2802--2810.

\bibitem[{Martin et~al.(2001)Martin, Fowlkes, Tal, and
  Malik}]{martin2001database}
Martin, D.; Fowlkes, C.; Tal, D.; and Malik, J. 2001.
\newblock A database of human segmented natural images and its application to
  evaluating segmentation algorithms and measuring ecological statistics.
\newblock In \emph{Proceedings Eighth IEEE International Conference on Computer
  Vision. ICCV 2001}, volume~2, 416--423. IEEE.

\bibitem[{Moran et~al.(2020)Moran, Schmidt, Zhong, and
  Coady}]{moran2020noisier2noise}
Moran, N.; Schmidt, D.; Zhong, Y.; and Coady, P. 2020.
\newblock Noisier2noise: Learning to denoise from unpaired noisy data.
\newblock In \emph{Proceedings of the IEEE/CVF Conference on Computer Vision
  and Pattern Recognition}, 12064--12072.

\bibitem[{Pl{\"o}tz and Roth(2018)}]{plotz2018neural}
Pl{\"o}tz, T.; and Roth, S. 2018.
\newblock Neural Nearest Neighbors Networks.
\newblock \emph{Advances in Neural Information Processing Systems}, 31:
  1087--1098.

\bibitem[{Quan et~al.(2020)Quan, Chen, Pang, and Ji}]{quan2020self2self}
Quan, Y.; Chen, M.; Pang, T.; and Ji, H. 2020.
\newblock Self2self with dropout: Learning self-supervised denoising from
  single image.
\newblock In \emph{Proceedings of the IEEE/CVF Conference on Computer Vision
  and Pattern Recognition}, 1890--1898.

\bibitem[{Ronneberger, Fischer, and Brox(2015)}]{ronneberger2015u}
Ronneberger, O.; Fischer, P.; and Brox, T. 2015.
\newblock U-net: Convolutional networks for biomedical image segmentation.
\newblock In \emph{International Conference on Medical image computing and
  computer-assisted intervention}, 234--241. Springer.

\bibitem[{Tai et~al.(2017)Tai, Yang, Liu, and Xu}]{tai2017memnet}
Tai, Y.; Yang, J.; Liu, X.; and Xu, C. 2017.
\newblock Memnet: A persistent memory network for image restoration.
\newblock In \emph{Proceedings of the IEEE international conference on computer
  vision}, 4539--4547.

\bibitem[{Ulyanov, Vedaldi, and Lempitsky(2018)}]{ulyanov2018deep}
Ulyanov, D.; Vedaldi, A.; and Lempitsky, V. 2018.
\newblock Deep image prior.
\newblock In \emph{Proceedings of the IEEE conference on computer vision and
  pattern recognition}, 9446--9454.

\bibitem[{Wu et~al.(2020)Wu, Liu, Cao, Ren, and Zuo}]{wu2020unpaired}
Wu, X.; Liu, M.; Cao, Y.; Ren, D.; and Zuo, W. 2020.
\newblock Unpaired learning of deep image denoising.
\newblock In \emph{European Conference on Computer Vision}, 352--368. Springer.

\bibitem[{Xie et~al.(2019)Xie, Liu, Li, Cheng, Zuo, Liu, Wen, and
  Ding}]{xie2019image}
Xie, C.; Liu, S.; Li, C.; Cheng, M.-M.; Zuo, W.; Liu, X.; Wen, S.; and Ding, E.
  2019.
\newblock Image inpainting with learnable bidirectional attention maps.
\newblock In \emph{Proceedings of the IEEE/CVF International Conference on
  Computer Vision}, 8858--8867.

\bibitem[{Xu et~al.(2020)Xu, Huang, Cheng, Liu, Zhu, Xu, and
  Shao}]{xu2020noisy}
Xu, J.; Huang, Y.; Cheng, M.-M.; Liu, L.; Zhu, F.; Xu, Z.; and Shao, L. 2020.
\newblock Noisy-as-clean: learning self-supervised denoising from corrupted
  image.
\newblock \emph{IEEE Transactions on Image Processing}, 29: 9316--9329.

\bibitem[{Xu et~al.(2018)Xu, Li, Liang, Zhang, and Zhang}]{xu2018real}
Xu, J.; Li, H.; Liang, Z.; Zhang, D.; and Zhang, L. 2018.
\newblock Real-world noisy image denoising: A new benchmark.
\newblock \emph{arXiv preprint arXiv:1804.02603}.

\bibitem[{Yu et~al.(2019)Yu, Lin, Yang, Shen, Lu, and Huang}]{yu2019free}
Yu, J.; Lin, Z.; Yang, J.; Shen, X.; Lu, X.; and Huang, T.~S. 2019.
\newblock Free-form image inpainting with gated convolution.
\newblock In \emph{Proceedings of the IEEE/CVF International Conference on
  Computer Vision}, 4471--4480.

\bibitem[{Zamir et~al.(2020)Zamir, Arora, Khan, Hayat, Khan, Yang, and
  Shao}]{zamir2020learning}
Zamir, S.~W.; Arora, A.; Khan, S.; Hayat, M.; Khan, F.~S.; Yang, M.-H.; and
  Shao, L. 2020.
\newblock Learning enriched features for real image restoration and
  enhancement.
\newblock In \emph{Computer Vision--ECCV 2020: 16th European Conference,
  Glasgow, UK, August 23--28, 2020, Proceedings, Part XXV 16}, 492--511.
  Springer.

\bibitem[{Zamir et~al.(2021)Zamir, Arora, Khan, Hayat, Khan, Yang, and
  Shao}]{zamir2021multi}
Zamir, S.~W.; Arora, A.; Khan, S.; Hayat, M.; Khan, F.~S.; Yang, M.-H.; and
  Shao, L. 2021.
\newblock Multi-stage progressive image restoration.
\newblock In \emph{Proceedings of the IEEE/CVF Conference on Computer Vision
  and Pattern Recognition}, 14821--14831.

\bibitem[{Zeyde, Elad, and Protter(2010)}]{zeyde2010single}
Zeyde, R.; Elad, M.; and Protter, M. 2010.
\newblock On single image scale-up using sparse-representations.
\newblock In \emph{International conference on curves and surfaces}, 711--730.
  Springer.

\bibitem[{Zhang et~al.(2017)Zhang, Zuo, Chen, Meng, and
  Zhang}]{zhang2017beyond}
Zhang, K.; Zuo, W.; Chen, Y.; Meng, D.; and Zhang, L. 2017.
\newblock Beyond a gaussian denoiser: Residual learning of deep cnn for image
  denoising.
\newblock \emph{IEEE transactions on image processing}, 26(7): 3142--3155.

\bibitem[{Zhang, Zuo, and Zhang(2018)}]{zhang2018ffdnet}
Zhang, K.; Zuo, W.; and Zhang, L. 2018.
\newblock FFDNet: Toward a fast and flexible solution for CNN-based image
  denoising.
\newblock \emph{IEEE Transactions on Image Processing}, 27(9): 4608--4622.

\bibitem[{Zhang et~al.(2011)Zhang, Wu, Buades, and Li}]{zhang2011color}
Zhang, L.; Wu, X.; Buades, A.; and Li, X. 2011.
\newblock Color demosaicking by local directional interpolation and nonlocal
  adaptive thresholding.
\newblock \emph{Journal of Electronic imaging}, 20(2): 023016.

\end{thebibliography}
\end{document}